\documentclass[12pt]{article}
\usepackage{citesort,fullpage,epsfig,psfrag,graphics,amsbsy,amssymb,amsmath
}
\usepackage{caption}
\newcommand{\beq}{\begin{equation}}
\newcommand{\eeq}{\end{equation}}
\newcommand{\bea}{\begin{eqnarray}}
\newcommand{\eea}{\end{eqnarray}}
\newcommand{\bfs}{\boldsymbol}

\newcommand{\Tr}{{\rm Tr}}
\newcommand{\be}{\begin{equation}}
\newcommand{\ee}{\end{equation}}
\newcommand{\bq}{\begin{eqnarray}}
\newcommand{\eq}{\end{eqnarray}}
\newcommand{\ket}[1]{|#1\rangle}

\def\math{\mathsurround=0pt }
\def\leftrightarrowfill{$\math \mathord\leftarrow \mkern-6mu 
 \cleaders\hbox{$\mkern-2mu \mathord- \mkern-2mu$}\hfill
 \mkern-6mu \mathord\rightarrow$}
\def\overleftrightarrow#1{\vbox{\ialign{##\crcr
     \leftrightarrowfill\crcr\noalign{\kern-1pt\nointerlineskip}
     $\hfil\displaystyle{#1}\hfil$\crcr}}}

\let\l=\lambda

 \def\bd{\begin{document}} \def\ed{\end{document}}
\def\ds{\documentstyle} \let\fr=\frac \let\bl=\bigl \let\br=\bigr
\let\Br=\Bigr \let\Bl=\Bigl
\let\bm=\bibitem
\let\na=\nabla
\let\pa=\partial \let\ov=\overline
\def\ft#1#2{{\textstyle{{\scriptstyle #1}\over {\scriptstyle #2}}}}
\def\fft#1#2{{#1 \over #2}}
\def\vp{\varphi}
\def\sst#1{{\scriptscriptstyle #1}}
\def\oneone{\rlap 1\mkern4mu{\rm l}}
\def\td{\tilde}
\def\wtd{\widetilde}
\def\dalemb#1#2{{\vbox{\hrule height .#2pt
        \hbox{\vrule width.#2pt height#1pt \kern#1pt
                \vrule width.#2pt}
        \hrule height.#2pt}}}
\def\square{\mathord{\dalemb{6.8}{7}\hbox{\hskip1pt}}}
\def\wtd{\widetilde}
\def\R{\rlap{\rm I}\mkern3mu{\rm R}}
\def\im{{\rm i}}
\def\tilg{\tilde{g}}
\def\tilF{\tilde{F}}
\def\tilA{\tilde{A}}
\def\varf{\varphi}
\def\tilf{\tilde{\phi}}
\def\tilh{\tilde{h}}
\def\rme{{\rm e}}
\def\ep{\epsilon}
\def\0{{(0)}}
\def\9{{(9)}}
\def\8{{(8)}}
\def\7{{(7)}}
\def\6{{(6)}}
\def\5{{(5)}}
\def\4{{(4)}}
\def\3{{(3)}}
\def\2{{(2)}}
\def\1{{(1)}}
\newcommand{\trace}{{\rm Tr}}
\newcommand{\ub}{\overline{U}}
\newcommand{\vb}{\overline{V}}
\newcommand{\uh}{\widehat{U}}
\newcommand{\vh}{\widehat{V}}
\newcommand{\ubh}{\overline{\widehat{U}}}
\newcommand{\vbh}{\overline{\widehat{V}}}
\newcommand{\lb}{\bar{\l}}
\newcommand{\Fb}{\overline{F}}
\newcommand{\Fh}{\widehat{F}}
\newcommand{\Fbh}{\overline{\widehat{F}}}
\newcommand{\Ab}{\overline{A}}
\newcommand{\Ah}{\widehat{A}}
\newcommand{\Abh}{\overline{\widehat{A}}}
\newcommand{\Gb}{\overline{G}}
\newcommand{\Gh}{\widehat{G}}
\newcommand{\Gbh}{\overline{\widehat{G}}}
\newcommand{\Pb}{\overline{P}}
\newcommand{\Ph}{\widehat{P}}
\newcommand{\Pbh}{\overline{\widehat{P}}}
\newcommand{\Qb}{\overline{Q}}
\newcommand{\Qh}{\widehat{Q}}
\newcommand{\Qbh}{\overline{\widehat{Q}}}
\newcommand{\Bb}{\overline{B}}
\newcommand{\Bh}{\widehat{B}}
\newcommand{\Bbh}{\overline{\widehat{B}}}
\newcommand{\fhns}{\hat{F}^{\rm (NS)}}
\newcommand{\fhrr}{\hat{F}^{\rm (RR)}}
\newcommand{\ahns}{\hat{A}^{\rm (NS)}}
\newcommand{\ahrr}{\hat{A}^{\rm (RR)}}
\newcommand{\hhrr}{\hat{H}^{\rm (RR)}}
\newcommand{\hchi}{\hat{\chi}}
\newcommand{\hphi}{\hat{\phi}}
\newcommand{\htau}{\hat{\tau}}
\newcommand{\cG}{{\cal G}}
\newcommand{\cGb}{\overline{{\cal G}}}
\newcommand{\cH}{{\cal H}}
\newcommand{\cP}{{\cal P}}
\newcommand{\cPb}{\overline{{\cal P}}}
\newcommand{\cQ}{{\cal Q}}
\newcommand{\cQb}{\overline{{\cal Q}}}
\newcommand{\cM}{{\cal M}}
\newcommand{\cN}{{\cal N}}
\newcommand{\cO}{{\cal O}}
\newcommand{\cD}{{\cal D}}
\newcommand{\cL}{{\cal L}}

\newcommand{\vpp}{\mbox{$\langle{\scriptstyle++}\rangle$}}
\newcommand{\vmp}{\mbox{$\langle{\scriptstyle-+}\rangle$}}
\newcommand{\vppp}{\mbox{$\langle{\scriptstyle+++}\rangle$}}
\newcommand{\vmpp}{\mbox{$\langle{\scriptstyle-++}\rangle$}}
\newcommand{\vpmp}{\mbox{$\langle{\scriptstyle+-+}\rangle$}}
\newcommand{\goesas}[1]{{}_{{\displaystyle\sim}\atop#1}}
\newcommand{\impliesas}[1]{{}_{{\displaystyle{\Longrightarrow}}\atop#1}}
\renewcommand{\thepage}{\arabic{page}}

\begin{document}
\setlength{\captionmargin}{36pt}
\begin{titlepage}
\begin{flushright}
\end{flushright}

\vskip 3cm
\begin{center}
\begin{large}
{\bf Protostring Scattering Amplitudes}
\end{large}
\vskip 2cm
{\large 
Charles B. Thorn
\footnote{E-mail  address: {\tt thorn@phys.ufl.edu}}
}
\vskip0.20cm
{\it Institute for Fundamental Theory,\\
Department of Physics, 
University of Florida
Gainesville FL 32611
}

\vskip24pt
\end{center}
\begin{abstract}
\noindent We calculate some tree level scattering amplitudes for
a generalization of the protostring, which is
a novel string model implied by the simplest string bit models. 
These bit models produce a lightcone worldsheet which supports
$s$ integer moded Grassmann fields. In the generalization we 
supplement this Grassmann worldsheet system with 
$d=24-s$ transverse coordinate worldsheet fields. 
The protostring corresponds to $s=24$ and the bosonic string
to $s=0$. The interaction vertex is a simple
overlap with no operator insertions at the break/join point. 
Assuming that $s$ is even
we calculate the multi-string scattering amplitudes 
by bosonizing the 
Grassmann fields, each pair equivalent to one compactified
bosonic field, and applying Mandelstam's interacting string formalism
to a system of $s/2$ compactified and $d$ uncompactified bosonic
worldsheet fields. We obtain all amplitudes for open strings
with no oscillator excitations and for closed strings with no 
oscillator excitations and zero winding number. We then study in detail
some simple special cases. Multi-string processes 
with maximal helicity violation
have much simplified amplitudes. We also specialize to 
general four string amplitudes and discuss their high energy behavior. 
Most of these models are not covariant under the full Lorentz
group $O(d+1,1)$. The exceptions are the bosonic string
whose Lorentz group is $O(25,1)$ and the protostring whose Lorentz
group is $O(1,1)$. The models in between only enjoy an $O(1,1)\times O(d)$
spacetime symmetry.
\end{abstract}

\vfill
\end{titlepage}
\section{Introduction}Recent studies of
string bit models \cite{gilest,thornsakh,bergmantsubit}
have motivated the serious consideration of some novel
string theories. String bits are hypothetical fundamental constituents 
of string. However just which string theories they describe depends
on their detailed structure. For example the superstring bit
\cite{bergmantsubit} underlying
IIB superstring is created by an operator ${\bar\phi}_{[a_1\cdots a_k]}({\bfs x})$,
where $a_i=1,\ldots,8$
are spinor indices, and $k=0,\ldots,8$. The vector variable ${\bfs x}$
gives the location of the bit in the eight dimensional transverse space of
lightcone quantized string theory \cite{goddardrt,goddardgrt}. 
These creation operators are also
$N\times N$ matrices in ``color'', the indices  of which we
suppress. In the large $N$ limit \cite{thooftlargen} the 
color singlet composites of $M$ 
string bits are closed chains which approximate continuous closed strings
for very large bit number $M$. On these long chains fluctuations
in the spinor degrees of freedom lead to the eight Grassmann worldsheet 
fields $\theta^a_{R,L}(\sigma)$ and fluctuations in the coordinates 
lead to the eight transverse coordinate worldsheet fields ${\bfs x}(\sigma)$.
In this interpretation a longitudinal
coordinate $x^-$ arises as the conjugate to a longitudinal
momentum identified with bit number $P^+\equiv Mm$.
Each string bit carries one unit $m$ of $P^+$. 
In this way holography \cite{thoofthologram} 
was realized by these models in its narrowest
sense: formulating $d+1$ dimensional physics in one dimension less.
These early string bit models, which naturally describe
type IIB superstring theory, inspired other matrix model proposals,
including the matrix model of $M$ theory \cite{banksfss} 
and the matrix string model
of type IIA superstring theory \cite{dvvmatrixstring}.

However, since the strings are simply large bit number composites,
the same string theory can arise from a string bit with considerably less
structure. A possibility proposed in \cite{sunthorn,chensun,thornspace,thornprotobits} 
is to replace each transverse coordinate ${x}^k$ of a string bit with
a two valued index whose fluctuations on long chains of bits simulate
the transverse space. In such string bit models all of space,
 both longitudinal $x^-$
{\it and} transverse ${\bfs x}$, are initially absent but emerge dynamically
in describing the physics of composites of string bits. 
But the string bit concept can be realized in more general ways
not necessarily tied to the known string models. Nonetheless certain
general constraints should still apply. In particular the constraint that
$1/N$ corrections lead to finite scattering amplitudes typically constrains
the size of the worldsheet field system. This is just the critical
dimension constraint in bosonic string ($D=26$) and superstring ($D=10$) theory.
In the latter case supersymmetry further links the spinor dimensionality
to the coordinate dimensionality.

In \cite{sunthorn,chensun,thornspace,thornprotobits} 
we studied, initially as a warm up
exercise, the simplest string bit model with only spinor degrees of freedom.
The bit creation operator carries only the spinor indices:
 ${\bar\phi}_{a_1\cdots a_k}$. Here we let $k=0,\ldots, s$
and $a_i=1,\ldots,s$, with the integer $s$ to be determined. 
These operators are bosonic if $k$ is even
and fermionic if $k$ is odd. They are also antisymmetric under the
interchange of any pair of the $a_i$. As explained in \cite{thornprotobits},
the large $N$ limit of these models predicts a worldsheet system
of $s$ left-right pairs of Grassmann worldsheet fields. Also the
$1/N$ corrections are described in string language as a simple overlap
of three string states, without the operator prefactor
familiar in the superstring vertex. It was found that this overlap
amplitude is finite in the continuum limit only if $s=24$. 
This is a novel string theory. In a sense it is more like the bosonic string
than the superstring because of the absence of prefactors in the
three string vertex. On the other hand the degree of freedom count of
worldsheet fields matches that of the superstring: bosonize
16 of the Grassmann dimensions to match the 8 transverse coordinates
and one is left with the 8 Grassmann dimensions of the superstring!
Unlike the bosonic string which has tachyons this emergent string,
which moves in one emergent space dimension $x^-$, has none: in
fact there is a mass gap. We call this string a protostring, since
it has the primitive simplicity (but not the instability) of the bosonic
string, and it also evokes intimations of superstring.
For this reason, we believe that the protostring is well worth
serious study in its own right.

The purpose of this article is to explore
the protostring's  physical properties further by 
determining its scattering amplitudes. 
We stress that the calculations we shall do
are pure string theory calculations: the fact that the protostring
was the outcome of a simple string bit model will play no role\footnote{
However, the underlying string bit physics may justify the use
of analytic continuation to define the integral representations 
of amplitudes which typically diverge for physical values
of the momenta.}. We only need
to faithfully apply Mandelstam's interacting lightcone string
formalism \cite{mandelstamlc,mandelstamnsr,mandelstamdet}.

The protostring can be succinctly characterized as the string model in which 
each of the 24 transverse coordinates
of the lightcone bosonic string is replaced by a spinor valued integer moded 
 Grassmann worldsheet field.
Here we shall consider a slight
generalization of the protostring in which only $s$ bosonic dimensions are 
so replaced,
with the remaining $d=24-s$ left as transverse coordinates. 
In all these models the
string interaction is a simple overlap without operator insertions
at the join/break point. The condition $s+d=24$ ensures the finite
continuum limit of the string bit overlap. 
With the continuum scattering
amplitude written in the form ${\cal M}\prod_k|p_k^+|^{-1/2}$, this finiteness
condition means that ${\cal M}$ is invariant under the scale transformation
$p_k^+\to\lambda p_k^+$. In other words, invariance under
the subgroup $SO(1,1)$ of the Lorentz group in $d+1$ space dimensions 
$SO(d+1,1)$ is maintained. For the protostring ($s=24$ or $d=0$), 
this is the entire Lorentz group, but for $s<24$,  with the exception of the
bosonic string ($s=0$), the Lorentz group is broken
to $SO(1,1)\times SO(d)$.

In the next Section 2 we explain the process of bosonization needed in the
rest of the paper. Then in Section 3 we apply Mandelstam's interacting
string formalism to the bosonized generalized protostring 
and calculate scattering
amplitudes for any number of external strings in states 
with zero winding number and no
oscillator excitations. Section 4 discusses amplitudes in special simplifying
circumstances. Section 5 analyzes high energy scattering in the
2 to 2 case. Concluding comments are in Section 6. An appendix reviews
the necessary measure calculations needed for the processes discussed
in the main text.
\section{Bosonization}
Hereafter we assume that $s$ is even, 
which allows us to approach the calculation of
scattering amplitudes by bosonizing each pair of the $s$ Grassmann dimensions.
Then we shall apply Mandelstam's lightcone interacting string formalism
for the bosonic string \cite{mandelstamlc} to calculate the
scattering amplitudes. When bosonized, the Grassmann system
is equivalent to $s/2$ compactified boson worldsheet fields $\phi^a$ 
in which the Kaluza-Klein (KK) momenta ${\bfs\pi}$ assume 
half odd integer multiples 
of the inverse compactification radius, which is fixed by the nature
of the Grassmann system.

 The bosonization procedure works only in the
continuum limit, $M\to\infty$. At finite $M$, in the notation of 
\cite{thornprotobits},
\bea
S_k&=&\frac{B_0}{\sqrt{M}}+\frac{1}{\sqrt{M}}\sum_{n=1}^{M-1}\left(F_ne^{-i\omega_n t}\cos\frac{\pi n}{2M}+{\bar F}_ne^{+i\omega_n t}\sin\frac{\pi n}{2M}\right)e^{2\pi ikn/M}\\
{\tilde S}_k&=&\frac{B_0}{\sqrt{M}}-i\frac{1}{\sqrt{M}}\sum_{n=1}^{M-1}\left(F_ne^{-i\omega_n t}\sin\frac{\pi n}{2M}-{\bar F}_ne^{+i\omega_n t}\cos\frac{\pi n}{2M}\right) e^{2\pi ikn/M}\\
\{F_n,{\bar F}_m\}&=&2\delta_{m+n,M},\qquad \{B_0,{\tilde B}_0\}=0,
\qquad B_0^2={\tilde B}_0^2=1
,\qquad \omega_n=
\frac{2T_0}{m}\sin\frac{n\pi}{M}
\eea
In the continuum limit $M\to\infty$ with $mM=P^+$ fixed, finite energy excitations have either $n$ or $M-n$
finite. The worldsheet coordinate $\sigma\sim km$ and the above formulas read
\bea
\frac{S_k}{\sqrt{m}}&\sim&\frac{B_0}{\sqrt{P^+}}+\sqrt{\frac{2}{{P^+}}}
\sum_{n=1}^{\infty}\left(f_ne^{-2i\pi n(T_0 t-\sigma)/P^+}+{f}_{n}^\dagger
e^{+2i\pi n(T_0 t-\sigma)/P^+}\right)\\
\frac{{\tilde S}_k}{\sqrt{m}}&\sim&\frac{{\tilde B}_0}{\sqrt{P^+}}
+\sqrt{\frac{2}{{P^+}}}
\sum_{n=1}^{\infty}\left({\tilde f}_{n}e^{-2i\pi n(T_0 t+\sigma)/P^+}
+{\tilde f}_{n}^\dagger e^{+2i\pi n(T_0 t+\sigma)/P^+}\right)\\
{f}_n&\equiv& \frac{F_{n}}{\sqrt{2}},\qquad{\tilde f}_n\equiv -i\frac{F_{M-n}}{\sqrt{2}},\qquad
\{f_n,{f}^\dagger_m\}=\{{\tilde f}_n,{\tilde f}^\dagger_m\}=\delta_{mn},
\eea
We see that in the continuum limit $S$ describes right moving and
${\tilde S}$ describes left-moving waves along the string. 

Labelling a pair of such Grassmann variables 1,2, the bosonization
formula for right-moving waves (see, for example Appendix A in
\cite{thornsft}) is
\bea
a_n&=&\frac{iB_0^1}{\sqrt{2}}f^2_n+f^1_n\frac{iB_0^2}{\sqrt{2}}
+i\sum_{k=1}^{n-1} f^1_kf^2_{n-k}+i\sum_{k=1}^\infty\left(
f^{1\dagger}_kf^2_{n+k}+f^1_{n+k}f^{2\dagger}_k\right), \qquad n>0\\
a_0&=& \frac{i}{2}B_0^1B_0^2+i\sum_{k=1}^\infty\left(f^{1\dagger}_kf^2_k
-f^{2\dagger}_kf^1_k\right),\qquad a_{-n}\equiv a_n^\dagger\\
{}[a_n,a_m]&=&n\delta_{n,-m}
\eea 
and similar formulas with tilde's for the left-moving waves ${\tilde a}_n$.
The square of the zero mode part of $a_0$ or ${\tilde a}_0$ is $1/4$, 
showing that the
lowest energy state has values $\pm1/2$ for $a_0$ and ${\tilde a}_0$.
The sum $a_0+{\tilde a}_0$ and difference $a_0-{\tilde a}_0$ have
the interpretation of KK momentum and winding number respectively.
We see that they have opposite parity: if one is
even the other is odd and vice versa. We shall be calculating
scattering amplitudes for zero winding number strings ($a_0={\tilde a}_0$) 
in which case
the KK momentum is an odd integer multiple of some scale.

It is sometimes convenient to express bosonization in terms of the
fermion operators of definite helicity, i.e. eigenoperators of $a_0$: 
\bea
b_0&=&\frac{1}{2}(B_0^1+iB_0^2),\qquad b_n=\frac{1}{\sqrt{2}}
(f_n^1+if_n^2),\qquad d_n=\frac{1}{\sqrt{2}}(f_n^1-if_n^2)\\
\{b_n,b_n^\dagger\}&=&\{d_n,d_n^\dagger\}\ =\ \{b_0,b_0^\dagger\}\ =\ 1,
\eea
all other anticommutators vanishing. In terms of these the boson operators are
\bea
a_n&=&d_nb_0+\sum_{k=1}^{n-1}d_kb_{n-k}+
b^\dagger_0b_n+\sum_{k=1}^\infty\left(b_k^\dagger b_{n+k}
-d_k^\dagger d_{n+k}\right),\qquad n>0\\
a_0&=&-\frac{1}{2}+b_0^\dagger b_0+\sum_{k=1}^{\infty}\left(b^\dagger_kb_k-d^\dagger_kd_k\right),\qquad a_{-n}=a_n^\dagger
\eea
In this article we discuss scattering amplitudes for external
strings in states with zero winding number and no oscillator excitations, which means they
are in states annihilated by $a_n,{\tilde a}_n$, $n>0$, 
but for which $a_0={\tilde a}_0$ can have
any allowed value. Since $[a_0,a_n]=[a_0,{\tilde a}_n]=0$, 
these states are the lowest
energy states with the given value of helicity. Define $\ket{0}$ to be
the lowest energy helicity $-1/2$ state, which means that
$(b_0,b_n,d_n)\ket{0}=0$. States with helicity $a_0=k-1/2$ are obtained
by applying $n$ $d^\dagger$'s and $m$ $b^\dagger$'s, with $m-n=k$, to
the state $\ket{0}$. Clearly the lowest energy states of definite helicity
are
\bea
&&d^\dagger_kd^\dagger_{k-1}\cdots d^\dagger_1\ket{0},\qquad a_0=-k-1/2,\quad
k=0,1,2,\ldots\\
&&b^\dagger_{k-1}b^\dagger_{k-2}\cdots b^\dagger_0\ket{0},\qquad a_0=k-1/2,\quad
k=1,2,\ldots
\eea
In Fermi language the energy of a state is the total mode number plus
$2/24$. The total mode numbers of these lowest energy states in each helicity
sector are $\sum_{l=1}^kl=k(k+1)/2$ and $\sum_{l=1}^{k-1}l=k(k-1)/2$. In
both cases the mode number is $a_0^2/2-1/8$. Adding on $2/24$, shows that
the energy of these states is precisely $a_0^2/2-1/24$, as the quantization
of the bosonic string requires.
\section{Scattering Amplitudes}
\subsection{Worldsheet path integral}
As we have seen, for strings of even winding number
there is a minimum nonzero
momentum magnitude which we call $\gamma$. Then the KK momenta
assume odd integer multiples of $\gamma$, ${\bfs\pi}=\pm(2k+1){\bfs\gamma}$,
$k=0,1,2,\ldots$.
Then the worldsheet path integral will be over $n_b=d+s/2=24-s/2$
bosonic fields.
In bosonized language, a nonzero three vertex will require an
insertion of the form $e^{\pm i\gamma \phi(\rho)}$ for each
$\phi^a$ at the join/break point. 
In other words
the KK momentum $\pi^k$ will not be conserved, with violation of up to
$\pm(N-2)\gamma$ for the $N$ point function.
Also, for hermiticity reasons, we shall require a hermitian 
linear combination of the insertion factors
with $\pm$ in the exponent, for example we can take the
insertion to be $e^{i\gamma\phi}+e^{-i\gamma\phi}=2\cos(\gamma\phi)$. 
We call the $d=24-s$ coordinate
fields $x^k$, the momenta ${\bfs p}$  of which will be continuous
and exactly conserved.
Thus in the evaluation of the scattering
amplitude we will allow a different (binary for each of the $s/2$ 
bosonized Grassmann dimensions) choice at each vertex: we
will write the insertion at the $r$th vertex as $
\prod_a\left(e^{i\gamma\phi_a(\rho_r)}+e^{-i\gamma\phi_a(\rho_r)}\right)$.
Then after expanding the vertex factors one has a collection 
of terms with a single exponential $e^{i\gamma^a_r\phi(\rho_r)}$
with $\gamma^a_r=\pm\gamma$. Then the momentum conservation law
following from Neumann boundary conditions for each such term
will take the form
\bea
\sum_{k=1}^N {\bfs\pi}_k+\sum_{r=1}^{N-2}{\bfs\gamma}_r=0,\qquad
\sum_{k=1}^N {\bfs p}_k=0,
\eea
for the bosonized Grassmann fields and regular coordinate fields
respectively.

It will be convenient to collect the ${\bfs p}_k,{\bfs\pi}_k$ of
the $k$th string participating in the scattering process
together in an
$n_b$ component transverse momentum ${\bfs P}_k=({\bfs p}_k,{\bfs\pi}_k)$.
We also denote the ``Minkowski'' $(n_b+2)$-vector with upper case Roman type,
$P_k=(p_k^-,p_k^+,{\bfs P}_k)$, and the Minkowski $(d+2)$-vector
with lower case Roman type, $p_k=(p_k^-,p_k^+,{\bfs p}_k)$.
The interacting string formalism 
automatically imposes the mass shell condition for an unexcited
string:
\bea
\alpha^\prime P\cdot P&=&
\begin{cases}
\displaystyle{\frac{n_b}{24}=1-\frac{s}{48}}&\qquad  {\rm Open\quad string}\\
&\\
\displaystyle{\frac{n_b}{6}=4-\frac{s}{12}}&\qquad {\rm Closed\quad string}
\end{cases}
\eea
These conditions reduce to the familiar $1$ and $4$ for the
bosonic string ($s=0$). 
Hereafter we shall choose units so that $\alpha^\prime=(2\pi T_0)^{-1}=1$.
Note that since ${\bfs\pi}^2\geq s\gamma^2/2$, the ``ordinary'' momentum
$p$ satisfies $p\cdot p\leq 1-s(1+24\gamma^2)/48$ or $p\cdot p
\leq 4-s((1+6\gamma^2)/12$ respectively. As we shall see, $\gamma^2=1/8,1/2$
respectively, so the theories with $s>12$ have a mass gap.

The contribution of each field $\phi^a$ to the Boltzmann factor
in the worldsheet path integrand for the $N$ string scattering amplitude
is
\bea
B(\phi)&=&\exp\left\{-\frac{1}{4\pi}\int d^2\rho (\nabla\phi)^2
+i\sum_r \gamma_r\phi(\rho(x_r))
+i\sum_k \frac{\pi_k}{2\pi|p_k^+|}\int d\sigma_k \phi(\sigma_k,\tau_k)\right\}
\eea
The contribution of each component of ${\bfs x}$ is similar,
 except that for them there
are no $\gamma$ terms, and the external momenta ${\bfs p}_k$ are continuous. 
The last term converts the description of the
 initial and final strings from
coordinate space to momentum space. Because we assume no
oscillator excitations of the $N$ strings, it suffices to
specify a constant
momentum density on each string at initial and final times. The
$\sigma_k$ for the $k$th string ranges over an interval of $\sigma$
of length $2\pi|p^+_k|$.
Here, to conform with Mandelstam's conventions, 
we have chosen the worldsheet spatial
coordinate $\sigma$ of a single string so that on a
given external string $0<\sigma_k-\sigma_k^0<2\pi p_k^+\equiv \pi\alpha_k$.
We also introduce the complex worldsheet coordinate $\rho=\tau+i\sigma$.
In the complex plane the integral in the last term can be viewed as a closed
line integral,
\bea
i\sum_k \frac{\pi_k}{2\pi |p_k^+|}\int d\sigma_k \phi(\sigma_k,\tau_k)
&=& i\oint ds J(\rho)\phi(\rho)\eea
where $J(\rho)=0$ on any horizontal boundaries and $J(\rho)=\pi_k/(2\pi |p^+_k|)$
on the vertical boundaries where strings enter or leave the diagram.
Our convention is that the ${\bfs \pi}_k,{\bfs p}_k$ will always be taken 
as incoming
momenta, and $p^+_k,\alpha_k$ are positive for incoming strings and negative for
outgoing strings. With this convention the absolute value signs
in $|p_k^+|$ can be removed by reversing the $\sigma_k$ integral for
outgoing external strings as in \cite{mandelstamlc}.

We can extract the $\gamma_r,\pi_k, p_k$ dependence of the path integral
by completing the square in the usual way: Shift
$\phi\to\phi+c$ and choose $c$ to cancel the linear terms:
\bea
-\nabla^2 c&=&2\pi i\sum_r \gamma_r\delta(\rho-\rho(x_r)),\qquad
\partial_n c|_{\partial}=2\pi i J\\
\ln {B(\phi+c)\over B(\phi)|_{0}}&=&
+{i\over2}\sum_s\gamma_s c(x_s)+{i\over2}
\sum_k{\pi_k\over 2\pi |p^+_k|}\int d\sigma_k c(\sigma_k,\tau_k)
\eea
The answer can be expressed in terms of the Neumann function
\bea
-\nabla^2N(\rho,\rho^\prime)=-2\pi\delta(\rho-\rho^\prime),\qquad
\partial_n N(\rho,\rho^\prime)|_{\rho\in\partial}=f(\rho)
\eea
Then applying Green's theorem we have
\bea
c(\rho^\prime)&=&-i\sum_r\gamma_r N(\rho(x_r),\rho^\prime)
-i\sum_{k}{\pi_k\over 2\pi |p^+_k|}\int d\sigma_kN(\rho,\rho^\prime)
\nonumber\\
&&+{1\over2\pi}\int d\sigma (cf)\bigg|^{\tau_f}_{\tau_i}
\eea
The last term, independent of $\rho^\prime$ drops out of $\ln B/B_0$
by momentum conservation:
\bea
\ln {B(\phi+c)\over B(\phi)|_{0}}&=&{1\over2}\sum_{r,s}\gamma_r\gamma_s
N(\rho(x_r),\rho(x_s))+\sum_{r,k}\gamma_r{\pi_k\over2\pi |p^+_k|}
\int d\sigma_k N(\rho(x_r),\rho_k)\nonumber\\
&&+{1\over2}\sum_{kl}
\int d\sigma_k d\sigma_l{\pi_k \pi_l\over4\pi^2|p^+_kp^+_l|}N(\rho_k,\rho_l)
\eea
The Neumann function for the lightcone worldsheet $\rho$ can be related to
that for the whole or half complex plane by the conformal mapping
$\rho(z)$ \cite{mandelstamlc} reviewed in
appendix A. For closed string amplitudes we map from the whole plane,
for which the Neumann function is
\bea
N(z,z^\prime)=\ln|z-z^\prime| .
\eea
In this case 
\bea
\ln {B(\phi+c)\over B(\phi)|_{0}}&=&\frac{1}{2}\sum_{r\neq s}\gamma_r\gamma_s
\ln|x_r-x_s|+\sum_{r,k}\gamma_r \pi_k\ln|x_r-Z_k|
+\frac{1}{2}\sum_{k\neq l}\pi_k \pi_l\ln|Z_k-Z_l|\nonumber\\
&&+\frac{\gamma^2}{2}\sum_r\ln|x_r-x_r| +{1\over2}\sum_{k}
\int d\sigma_k d\sigma^\prime_k{\pi^2_k\over4\pi^2p^{+2}_k}
N(\rho_k,\rho^\prime_k),\quad {\rm Closed}
\label{closedbfactor}
\eea
where $\rho(x_r)$ are the break/join points of the lightcone diagram
and $\rho(Z_k)$ are the locations of the incoming and outgoing strings. 

In contrast, for open string amplitudes we need the Neumann function for the upper half plane, 
\bea
N(z,z^\prime)=\ln|z-z^\prime|+\ln|z-z^{\prime*}|\to 2\ln|z-z^\prime|
\eea
when one or both $z$'s are on the real axis. Then, we find
\bea
\ln {B(\phi+c)\over B(\phi)|_{0}}&=&\sum_{r\neq s}\gamma_r\gamma_s
\ln|x_r-x_s|+2\sum_{r,k}\gamma_r \pi_k\ln|x_r-Z_k|
+\sum_{k\neq l}\pi_k \pi_l\ln|Z_k-Z_l|\nonumber\\
&&+\gamma^2\sum_r\ln|x_r-x_r| +{1\over2}\sum_{k}
\int d\sigma_k d\sigma^\prime_k{\pi^2_k\over4\pi^2p^{+2}_k}
N(\rho_k,\rho^\prime_k),\quad {\rm Open}
\label{openbfactor}
\eea
Note that $Z_N$, which we have set to $\infty$, appears on the
right side of (\ref{closedbfactor}) or (\ref{openbfactor}) in the combination
\bea
2\pi_N\left(\sum_r\gamma_r+\sum_{k=1}^{N-1}\pi_k\right)\ln Z_N
=-2\pi_N^2\ln Z_N
\eea
so the terms involving $Z_N$  will
enter, just as the ordinary transverse dimensions, into the terms
that implement the mass shell condition and wave function on the $N$th leg
\cite{mandelstamlc}.
\subsection{Self contractions}
The self-contractions on the last line of (\ref{closedbfactor}) or (\ref{openbfactor})
 need further discussion.
Those in the last term are of the same form for all $d+s/2$ 
bosonic fields
and combined give the mass shell condition. Consulting the mapping function 
(\ref{rhoofzo}) or (\ref{rhoofzc}), for $z$ near $Z_k$, we find
\bea
z-Z_k&\sim&e^\rho\prod_{l\neq k}(Z_k-Z_l)^{-\alpha_l/\alpha_k}\\
\ln|z-z^\prime|&\sim&\ln|e^\rho-e^{\rho^\prime}|+\sum_{l\neq k}
\ln|Z_k-Z_l|^{-\alpha_l/\alpha_k}
\eea
The contribution of the first term to the $\phi$ insertion is part
of the external wave function that is to be amputated, whereas
the second term provides the factors
\bea
\prod_{k\neq l}|Z_k-Z_l|^{-\pi_k^2\alpha_l/\alpha_k}
=\prod_{k<l}|Z_k-Z_l|^{-\pi_k^2\alpha_l/\alpha_k-\pi_l^2\alpha_k/\alpha_l}
\eea
for the open string, or
\bea
\prod_{k\neq l}|Z_k-Z_l|^{-\pi_k^2\alpha_l/(2\alpha_k)}
=\prod_{k<l}|Z_k-Z_l|^{-\pi_k^2\alpha_l/(2\alpha_k)-\pi_l^2\alpha_k/(2\alpha_l)}
\eea 
for the closed string.
The exponents in these factors contribute the $\pi_k$ dependent part of 
$p_k^-$ in the scalar product
\bea
P_k\cdot P_l&=&{\bfs p}_k\cdot{\bfs p}_l+{\bfs\pi}_k\cdot{\bfs \pi}_l
-p_k^+p_l^--p_k^-p_l^+=p_k\cdot p_l+{\bfs\pi}_k\cdot{\bfs \pi}_l,
\eea
which are the eventual powers in the Koba-Nielsen
factors $\prod_{k< l}|Z_k-Z_l|^{2P_k\cdot P_l}$ or $\prod_{k< l}|Z_k-Z_l|^{P_k\cdot P_l}$
in the open and closed string integrands respectively.
The lightcone
mass shell condition for an open or closed string with no
oscillator excitations reads
\bea
p_k\cdot p_k&=&P_k\cdot P_k-{\bfs\pi}_k\cdot{\bfs\pi_k}
={\bfs p}_k^2-2p^+_kp^-_k=1-\frac{s}{48}-{\bfs\pi}_k^2
\leq1-\frac{s}{48}-\frac{s}{2}\gamma^2\nonumber\\
p_k\cdot p_k&=&4-\frac{s}{12}-{\bfs\pi}_k^2
\leq4-\frac{s}{12}-\frac{s}{2}\gamma^2\nonumber\eea
The left side of this equation is the Minkowski scalar product 
 $p_{k\mu}p^\mu_k$ in $d+2=26-s$ space-time dimensions.
The inequality follows because each component of ${\bfs\pi}^k$ 
is an odd multiple of $\gamma$. 

Next we give our interpretation of the self contractions at
the interaction points. Infinities in these contractions can be
absorbed into the coupling constant, provided they are independent of the
geometry of the worldsheet. Since the lightcone worldsheet is
the fundamental starting point, we should set any regulator cutoffs 
in the $\rho$ coordinate. The same mapping function is used for open and closed
strings, except for a factor of two. Let us examine $\rho(z)$ near $z=x_r$,
the break/join point where $d\rho/dz=0$. For the open string case,
we expand
\bea
\rho(z)&=& \rho(x_r) +\frac{1}{2}{d^2\rho\over dz^2}\bigg|_{z=x_r}
(z-x_r)^2+O((z-x_r)^3)
\label{rhonearx}
\eea
Then use (\ref{rho2primeo}) for the second derivative
and  solve (\ref{rhonearx})
\bea
z-x_r&\approx&\sqrt{2}\sqrt{\rho-\rho(x_r)}
\left[{\prod_{k}(x_r-z_k)\over-\alpha_N\prod_{s\neq r}(x_r-x_s)}
\right]^{1/2}\\
|z(\rho)-z(\rho^\prime)|&\approx&|\sqrt{2(\rho-\rho(x_r))}
-\sqrt{2(\rho^\prime-\rho(x_r))}
|{\prod_{k}\sqrt{|x_r-z_k|}\over\sqrt{|\alpha_N|}
\prod_{s\neq r}\sqrt{|x_r-x_s|}}
\eea
We then interpret the self contraction as
\bea
\gamma^2\sum_r\ln|x_r-x_r|&\to&{\gamma^2\over2}\left[(N-2)\ln(2\delta)
+\ln{\prod_{r,k}
{|x_r-Z_k|}\over{|\alpha_N|^{N-2}}
\prod_{s\neq r}{|x_r-x_s|}}\right]\\
&&\hskip-2.5cm\to{\gamma^2\over2}\left[(N-2)\ln(2\delta)+\ln{|\alpha_N|\over
\prod_{k=1}^{N-1}|\alpha_k|}+\ln{\prod_{r,k}
{|x_r-Z_k|^2}\over
\prod_{s\neq r}{|x_r-x_s|}\prod_{k\neq l}|Z_k-Z_l|}\right]
\eea
where we have let $\delta$ be a measure of the cutoff regularization on the
lightcone world sheet.
We used the identity (\ref{identity2}) to arrive at 
the last line, which gives our interpretation of the self contractions at
the interaction points.
\subsection{The amplitudes}
Having taken care of the terms involving $Z_N$ and the self contractions
at the external states, 
what remains of the contribution of the boundary data to the
open string path integrand including all $s/2$ 
compactified bosonic dimensions, and $d=24-s$ ordinary uncompactified
bosonic dimensions is:
\bea
I_O&=&(2\delta)^{s(N-2)\gamma^2/4}\left[{\prod_{r\neq s}
|x_r-x_s|^{{\bfs\gamma}_r\cdot{\bfs\gamma}_s
-s\gamma^2/4}
\over\prod_{r,k<N}|x_r-Z_k|^{-2{\bfs\pi}_k\cdot{\bfs\gamma}_r-s\gamma^2/2}}
\right]\left[
{|\alpha_N|\over\prod_{k=1}^{N-1}|\alpha_k|}\right]^{s\gamma^2/4}\nonumber\\
&&\times\prod_{k<l}|Z_k-Z_l|^{2{\bfs P}_k\cdot{\bfs P}_l-s\gamma^2/2-{\bfs P}_k^2\alpha_l/\alpha_k-{\bfs P}_l^2\alpha_k/\alpha_l}
\eea
For the closed string, the factor of 2 in the mapping function leads
to some minor modifications:
\bea
I_C&=&(4\delta)^{s(N-2)\gamma^2/8}
{\prod_{r\neq s}|x_r-x_s|^{{\bfs\gamma}_r\cdot{\bfs\gamma}_s/2
-s\gamma^2/8}
\over\prod_{r,k<N}|x_r-Z_k|^{-{\bfs\pi}_k\cdot{\bfs\gamma}_r-s\gamma^2/4}}
\left[
{|\alpha_N|\over\prod_{k=1}^{N-1}|\alpha_k|}\right]^{s\gamma^2/8}\nonumber\\
&&\times\prod_{k<l}|Z_k-Z_l|^{{\bfs P}_k\cdot{\bfs P}_l-s\gamma^2/4
-{\bfs P}_k^2\alpha_l/(2\alpha_k)-{\bfs P}_l^2\alpha_k/(2\alpha_l)}\eea
The boundary data for one of the ${\bfs x}$ fields contributes the
same but with no $\gamma$ terms.

 The Jacobian and determinant factors (see (\ref{jacobdetopen2})
and (\ref{jacobdetclosed2}) in the appendix)
 will have $n_b=d+s/2=24-s/2$. Combining these with the $I$ 
factors gives for the open string, 
\bea
&&\hskip-1.5cm \left[I_O\left|{\partial T\over\partial Z}\right| 
{\det}^{-(24-s/2)/2}(-\nabla^2)\right]_{\rm open}\nonumber\\
&=&\prod_{k=1}^N{1\over \sqrt{|
\alpha_k|}}\left[{\prod_{k<N}|\alpha_k|\over
|\alpha_N|}\right]^{s(1-8\gamma^2)/32}\left[{\prod_{r<t}|x_t-x_r|
\prod_{m<l}|Z_l-Z_m|
\over \prod_{l,r}|Z_l-x_r|}\right]^{s/48}\nonumber\\
&&(2\delta)^{s(N-2)\gamma^2/4}\left[{\prod_{r<s}|x_r-x_s|^{2{\bfs\gamma}_r\cdot{\bfs\gamma}_s
-s\gamma^2/2}
\prod_{k< l<N}|Z_k-Z_l|^{2P_k\cdot P_l-s\gamma^2/2}
\over\prod_{r,k<N}|x_r-Z_k|^{-2{\bfs \pi}_k{\bfs\gamma}_r-s\gamma^2/2}}\right]
\eea
In string bit models, the factor within the first set of square brackets would scale as $M^{N-2}$ where $M$ is the bit number. So 
a finite continuum limit requires $\gamma^2=1/8$, in which case
\bea
&&\hskip-1.5cm \left[I_O\left|{\partial T\over\partial Z}\right| 
{\det}^{-(24-s/2)/2}(-\nabla^2)\right]_{\rm open}\nonumber\\
&=&(2\delta)^{s(N-2)/32}\prod_{k=1}^N{1\over \sqrt{|
\alpha_k|}}\left[{\prod_{r<s}|x_r-x_s|^{2{\bfs\gamma}_r\cdot{\bfs\gamma}_s
-s/24}
\prod_{k< l<N}|Z_k-Z_l|^{2P_k\cdot P_l-s/24}
\over\prod_{r,k<N}|x_r-Z_k|^{-2{\bfs\pi}_k{\bfs\gamma}_r-s/24}}\right]
\label{openintegrand}
\eea
Applying parallel considerations to the closed string leads to
\bea
&&\hskip-1.5cm \left[I_C\left|{\partial T\over\partial Z}\right|^2 
{\det}^{-(24-s/2)/2}(-\nabla^2)\right]_{\rm closed}\nonumber\\
&=&\prod_{k=1}^N{1\over{|
\alpha_k|}}\left[{\prod_{k<N}|\alpha_k|\over
|\alpha_N|}\right]^{s/16-s\gamma^2/8}\left[{\prod_{r<t}|x_t-x_r|
\prod_{m<l}|Z_l-Z_m|
\over \prod_{l,r}|Z_l-x_r|}\right]^{s/24}\nonumber\\
&&(4\delta)^{s(N-2)\gamma^2/8}\left[{\prod_{r<s}|x_r-x_s|^{{\bfs\gamma}_r\cdot{\bfs\gamma}_s-s\gamma^2/4}
\prod_{k< l<N}|Z_k-Z_l|^{P_k\cdot P_l-s\gamma^2/4}
\over\prod_{r,k<N}|x_r-Z_k|^{-{\bfs\pi}_k\cdot{\bfs\gamma}_r-s\gamma^2/4}}\right]
\eea
and a smooth continuum limit in the closed case requires $\gamma^2=1/2$:
\bea
&&\hskip-1.5cm \left[I_C\left|{\partial T\over\partial Z}\right|^2 
{\det}^{-(24-s/2)/2}(-\nabla^2)\right]_{\rm closed}\nonumber\\
&=&(4\delta)^{s(N-2)/16}\prod_{k=1}^N{1\over{|
\alpha_k|}}\left[{\prod_{r<s}|x_r-x_s|^{{\bfs\gamma}_r\cdot{\bfs\gamma}_s
-s/12}
\prod_{k< l<N}|Z_k-Z_l|^{P_k\cdot P_l-s/12}
\over\prod_{r,k<N}|x_r-Z_k|^{-{\bfs\pi}_k\cdot{\bfs\gamma}_r-s/12}}\right]
\label{closedintegrand}
\eea
Notice how the power of $|Z_k-Z_l|$ combined in the simplification:
\bea
2{\bfs P}_k\cdot{\bfs P}_l-s/16+s/48-({\bfs P}_k^2-n_b/24)\alpha_l/\alpha_k
-({\bfs P}_l^2-n_b/24)\alpha_k/\alpha_l&=&2P_k\cdot P_l-s/24\nonumber\\
{\bfs P}_k\cdot{\bfs P}_l-s/8+s/24
-({\bfs P}_k^2-n_b/6)\alpha_l/(2\alpha_k)-({\bfs P}_l^2-n_b/6)\alpha_k/(2\alpha_l)&=&P_k\cdot P_l-s/12\nonumber\eea
for open and closed strings respectively.
In these formulas each component of ${\bfs\gamma_r}$ is 
$\pm\gamma=\pm1/(2\sqrt{2})$ for the open string and $\pm\gamma=\pm1/\sqrt{2}$
for the closed string. 
And of course each component of ${\bfs \pi}_k$ is an odd integer multiple of
$\gamma$.

The scattering amplitudes are obtained by integrating the expression
(\ref{openintegrand}) or (\ref{closedintegrand})
over the unfixed $Z_k$. In the case of the open string, the $Z_k$ are
on the real axis satisfying $Z_1=0<Z_2<\cdots<Z_{N-2}<Z_{N-1}=1$. In the
case of the closed string the $Z_k$ for $k=2,\ldots N-2$ are integrated
over the whole complex plane. In both cases $Z_1=0, Z_{N-1}=1, Z_N=\infty$.
We remind the reader that for physical values of the momenta the resulting
integrals are plagued with divergences. To handle these divergences, one
starts with (unphysical) values of the momenta for which the integrals
converge, and then one analytically continues to the physical values.
For open string amplitudes one can do this keeping the range of
the $Z$ integrations complete. But for closed string amplitudes one
is forced to divide the integration region up into cells, with
separate analytic continuations in each cell. 

\section{Scattering amplitudes in special cases}
\subsection{External strings of minimal mass}
The scattering amplitudes obtained in the previous section described
$N$ external strings, all with zero winding number
and no oscillator excitations. However
the compactified momenta ${\bfs\pi}_k$ could have components any odd
multiple of $\pm\gamma$.
To specialize to external strings of minimal mass, each component of
each  ${\bfs \pi}_k$
should be $\pm\gamma$. In that case the mass squared of each string
is $-p\cdot p=s/12-1$ for the open string, or $s/3-4$ for the closed string.
There are all together
$N+N-2=2(N-1)$ ${\bfs\gamma}$'s and ${\bfs\pi}$'s, so to satisfy the 
conservation law 
$N-1$ should have value $+\gamma$ and the remaining should have value
$-\gamma$. There are $2(N-1)\choose(N-1)$ ways to do this for each
of the $s/2$ compactified bosonic fields. 

A dramatic simplification occurs when there
is maximal helicity violation. For instance, choose all components of
the first $N-1$ ${\bfs \pi}_k$ to have the value $-\gamma$.
Then necessarily all components of ${\bfs \pi}_N$ and of
each ${\bfs\gamma_r}$  have the value $+\gamma$.  In this case 
${\bfs\gamma}_r\cdot{\bfs\gamma_s}={\bfs \pi}_k\cdot{\bfs \pi}_l
=-{\bfs\gamma}_r\cdot{\bfs \pi}_l=s\gamma^2/2$ for $k,l\neq N$ and 
${\bfs \pi}_N$
doesn't appear in the formula. Then for the open case ($\gamma^2=1/8$) the 
contribution to the integrand of the scattering amplitude is
\bea
&&\hskip-1.5cm I_O\left|{\partial T\over\partial Z}\right| 
{\det}^{-(24-s/2)/2}(-\nabla^2)_{\rm open}\nonumber\\
&\to&\epsilon^{s(N-2)/32}\prod_{k=1}^N{1\over \sqrt{|
\alpha_k|}}\left[{\prod_{r<s}|x_r-x_s|^{2}
\prod_{k< l<N}|Z_k-Z_l|^{2}
\over\prod_{r,k<N}|x_r-Z_k|^{2}}\right]^{s/24}|Z_k-Z_l|^{p_k\cdot p_l}\\
&\to&\epsilon^{s(N-2)/32}\prod_{k=1}^N{1\over \sqrt{|
\alpha_k|}}\frac{|
\alpha_N|^{s(N-1)/12}}{\prod_{k<N}|\alpha_k|^{s/12}}
\left[\frac{\prod_{r<s}|x_r-x_s|^{s/12}}
{\prod_{k< l<N}|Z_k-Z_l|^{s/12}}\right]|Z_k-Z_l|^{p_k\cdot p_l}\eea
where we made use of (\ref{identity2}) to arrive at the last line.
With this simple choice the scattering amplitude is then
\bea
A^{\rm Open}_N&=&g^{N-2}\prod_{k=1}^N{1\over \sqrt{|\alpha_k|}}
\frac{|\alpha_N|^{s(N-1)/12}}{\prod_{k<N}|\alpha_k|^{s/12}}
\int dZ_2\cdots dZ_{N-1}\nonumber\\
&&{\prod_{r<s}|x_r-x_s|^{s/12}}
{\prod_{k< l<N}|Z_k-Z_l|^{2p_k\cdot p_l-s/12}}
\label{equalgammao}
\eea
Making the same simplifications for the case of the closed string leads to
\bea
A^{\rm Closed}_N
&=&g^{2(N-2)}\prod_{k=1}^N{1\over{|
\alpha_k|}}\frac{|
\alpha_N|^{s(N-1)/6}}{\prod_{k<N}|\alpha_k|^{s/6}}
\int d^2Z_2\cdots d^2Z_{N-1}\nonumber\\
&&{\prod_{r<s}|x_r-x_s|^{s/6}}
{\prod_{k<l<N}|Z_k-Z_l|^{p_k\cdot p_l-s/6}}
\label{equalgammac}
\eea
If desired, one can replace $2p_k\cdot p_l$ by $(p_k+p_l)^2-2+s/6$ 
in the open case
and by $(p_k+p_l)^2-8+2s/3$ in the closed case.
Keep in mind that this simpler expression only applies for a very special
choice for the ${\bfs\pi}$'s and ${\bfs\gamma}$'s.
In particular, even for a particular set of 
the ${\bfs\pi}_k$, the full insertion factor at each vertex 
is $2\cos(\gamma\phi)$, which
is to be implemented by summing the amplitudes over each component of
each ${\bfs \gamma}_r$ assuming both possible values $\pm\gamma$.
\subsection{Three string amplitudes}
In the case $N=3$ the three $Z$'s are fixed at $0,1,\infty$. The 
relevant conformal map is
\bea
\rho&=&\alpha_1\ln z+\alpha_2\ln(z-1)
\eea
which determines $x=\alpha_2/(\alpha_1+\alpha_2)$ and 
$1-x=\alpha_1/(\alpha_1+\alpha_2)$. Then the open string amplitude 
(\ref{openintegrand}) for this case reduces to:
\bea
\frac{1}{\sqrt{|\alpha_1\alpha_2\alpha_{12}|}}
\left|\frac{\alpha_2}{\alpha_{12}}\right|^{2{\bfs\pi}_1\cdot{\bfs\gamma}+s/24}
\left|\frac{\alpha_1}{\alpha_{12}}\right|^{2{\bfs\pi}_2\cdot{\bfs\gamma}+s/24}
\eea
The corresponding three closed string vertex is
\bea
\frac{1}{{|\alpha_1\alpha_2\alpha_{12}|}}
\left|\frac{\alpha_2}{\alpha_{12}}\right|^{{\bfs\pi}_1\cdot{\bfs\gamma}+s/12}
\left|\frac{\alpha_1}{\alpha_{12}}\right|^{{\bfs\pi}_2\cdot{\bfs\gamma}+s/12}
\eea
These formulas are valid only if all three strings are on shell. Putting
each $p^-_k=({\bfs p}_k^2+m_k^2)/\alpha_k$ the on shell
condition is $p_1^-+p_2^-+p_3^-=0$, which leads to a quadratic
equation determining the variable $x=\alpha_2/\alpha_{12}$. The condition that
$x$ is real and $0<x<1$, which must hold if $\alpha_1$ and $\alpha_2$ have the same sign, requires that the mass of particle 3 is greater than the sum
of masses 1 and 2 or $m_3^2\geq (m_1+m_2)^2$. 
This is just the requirement that the decay of
particle 3 into particles 1 and 2 is energetically allowed.

The squared mass of a string in the unexcited states, which we are considering,
is given by ${\bfs\pi}^2+s/48-1$ for the open string and
by ${\bfs\pi}^2+s/12-4$ for the closed string. 
For example, the decay into equal 
masses requires, in the case of open strings,
\bea
({\bfs\pi}_1+{\bfs\pi}_2+\bfs\gamma)^2+\frac{s}{48}-1
\geq 4{\bfs\pi}_1^2+\frac{s}{12}-4
\eea
which is easily satisfied if ${\bfs\gamma},{\bfs\pi}_1,{\bfs\pi}_2$
are sufficiently aligned.

\subsection{Four string amplitudes}
First specialize the conformal mapping from $z$ to $\rho$ to four
external strings,
\bea
\rho&=&\alpha_1\ln z+\alpha_2\ln(z-Z)+\alpha_3\ln(z-1)
\eea
Then the two interaction points are determined by $d\rho/dz=0$,
which implies the quadratic equation
\bea
0&=&(\alpha_1+\alpha_2+\alpha_3)x^2+(-\alpha_1(Z+1)-\alpha_2-\alpha_3Z)x
+\alpha_1Z\nonumber
\eea
with solutions
\bea
x_\pm&=&\frac{\alpha_1(Z+1)+\alpha_2+\alpha_3Z\pm
\sqrt{(\alpha_1(Z+1)+\alpha_2+\alpha_3Z)^2+4\alpha_1\alpha_4Z}
}{2(\alpha_1+\alpha_2+\alpha_3)}
\eea
which lead to
\bea
\alpha_4^2|x_+-x_-|^2
&=&\alpha_{12}^2(1-Z)+\alpha_{23}^2Z
-\alpha_{13}^2Z(1-Z)\eea
where $\alpha_{kl}\equiv \alpha_k+\alpha_l$.

Since the $x_\pm$ enter the integrand of the scattering amplitude, their
behavior as $Z\to0,1$ which control the pole locations in the variables
$(p_1+p_2)^2$ and $(p_2+p_3)^2$ is relevant. We find
\bea
x_\pm&\sim&\begin{cases}\displaystyle{-{(\alpha_1+\alpha_2)}/{\alpha_4}
\choose Z\alpha_1/\alpha_{12}}&\qquad {\rm for}\quad Z\sim0\\
\displaystyle{-{\alpha_1}/{\alpha_4}
\choose1-(Z-1)\alpha_3/\alpha_{14}}&\qquad {\rm for}\quad Z\sim1
\end{cases}
\label{limitexes}
\eea  
As long as $\alpha_{12}$ and $\alpha_{23}$ are both non zero we only need to
pay attention to the factors $x_-$ and $Z-x_-$ when analyzing $Z\sim0$,
and to the factors $1-x_-$ and $Z-x_-$ when $Z\sim1$.

Putting $N=4$ in (\ref{equalgammao}) we find the four open string amplitude
in the special case $\gamma_r=\gamma=1/\sqrt{8}$, $\pi_k=-\gamma=-1/\sqrt{8}$ 
for $k<4$:
\bea
A_{4}^{\rm open}=g^{2}\prod_{k=1}^4\frac{1}{\sqrt{|
\alpha_k|}}\int_0^1 dZ
\frac{|\alpha_4|^{s/4}|x_2-x_1|^{s/12}}{|\alpha_1\alpha_2\alpha_3|^{s/12}}
Z^{(p_1+p_2)^2-2+s/12}(1-Z)^{(p_2+p_3)^2-2+s/12}
\eea
It is not hard to check that the pole singularities in
$(p_1+p_2)^2$ and $(p_2+p_3)^2$ are where they should be
as long as $\alpha_{12}$ and $\alpha_{23}$ are non zero. This is reasonable since
excluding these values of the $\alpha$'s guarantees that the 
dynamical singularities are all due to the long time propagation 
of protostring mass eigenstates.
If, for example, $\alpha_{23}=0$, $\alpha_4^2(x_2-x_1)^2\sim4\alpha_1\alpha_2
(1-Z)$ as $Z\to1$ so the poles in $(p_2+p_3)^2$ are shifted by an amount
$s/24$. When $\alpha_{23}=0$ these singularities are due to the collision
of the interaction points on the worldsheet and not the long time
propagation of a particle state. This nonuniformity of singularity
structure is absent for the bosonic and superstring because the
amplitude integrands then turn out to be independent of the $x_r$. 

The four closed string amplitude (\ref{equalgammac})
in the case $N=4$ and   $\gamma_r=\gamma=1/\sqrt{2}$, $\pi_k=-\gamma=-1/\sqrt{2}$
for $k<4$ is
\bea
A^{\rm Closed}_4
&=&g^{4}\prod_{k=1}^4{1\over{|
\alpha_k|}}\frac{|
\alpha_4|^{s/2}}{|\alpha_1\alpha_2\alpha_3|^{s/6}}
\int d^2Z\nonumber\\
&&{|x_2-x_1|^{s/6}}
|Z|^{(p_1+p_2)^2/2-4+s/6}|1-Z|^{(p_2+p_3)^2/2-4+s/6}
\eea
\subsection{Four protostring amplitude}
In the case of the protostring ($s=24$) space is only one dimensional.
This puts severe limits on the kinematics: scattering can be only
forward ($\alpha_{23}=0$) or backward ($\alpha_{13}=0$). To evaluate protostring
open string amplitudes, we keep $s<24$ as a regulator but 
restrict the kinematics
to 2 space-time dimensions. Thus we set all the ${\bfs p}_k=0$, but keep
the general $s$ mass values $m_k^2=(s-12)/12$. Then the 
Mandelstam invariants are
\bea
S&=&-(p_1+p_2)^2=\frac{\alpha_{12}^2}{\alpha_1\alpha_2}\frac{s-12}{12}\\
t&=&-(p_2+p_3)^2=\begin{cases}0&\qquad \alpha_{23}=0\\
\frac{s-12}{3}-S&\qquad \alpha_{13}=0\end{cases}\\
u&=&-(p_1+p_3)^2=\begin{cases}0&\qquad \alpha_{13}=0\\
\frac{s-12}{3}-S&\qquad \alpha_{23}=0\end{cases}
\eea
and we have
\bea
\alpha_4^2|x_+-x_-|^2&=&\begin{cases}
\alpha_{12}^2(1-Z)^2+4\alpha_{1}\alpha_2Z(1-Z)&\qquad \alpha_{23}=0\\
\alpha_{12}^2-4\alpha_1\alpha_2Z&\qquad \alpha_{13}=0
\end{cases}\eea
The maximal helicity violating
four point function of the preceding section becomes, for
forward scattering and setting $s=24$,
\bea
A_4^{\alpha_{23}=0}&=&g^{2}\prod_{k=1}^4{1\over \sqrt{|
\alpha_k|}}\frac{|\alpha_1|^{2}}{|\alpha_2|^4}
\int dZ\left[{\alpha_{12}^2(1-Z)
+4\alpha_1\alpha_2 Z}\right]Z^{-S}(1-Z)^{1}\nonumber\\
&=&g^{2}\prod_{k=1}^4{1\over \sqrt{|
\alpha_k|}}\frac{|\alpha_1|^{2}}{|\alpha_2|^4}\left[\alpha_{12}^2
\frac{\Gamma(-S+1)
\Gamma(3)}{\Gamma(-S+4)}
+4\alpha_1\alpha_2
\frac{\Gamma(-S+2)\Gamma(2)}{\Gamma(-S+4)}\right]
\nonumber\\
&=&\frac{g^{2}}{|
\alpha_1\alpha_2|}\frac{|\alpha_1|^{2}}{|\alpha_2|^4}
\left[\frac{2\alpha_{12}^2}{(1-S)(2-S)(3-S)}
+\frac{4\alpha_1\alpha_2}{(2-S)(3-S)}\right]
\eea
In doing these integrals, we begin with $S<0$ so that the integrals
converge. One can then continue $S$ to $S>4$ to reach physical 
scattering. 

For backward scattering, we temporarily keep $s\neq24$  
in the powers of $Z$ and $1-Z$.
\bea
A_4^{\alpha_{13}=0}&=&g^{2}\prod_{k=1}^4{1\over \sqrt{|
\alpha_k|}}\frac{1}{|\alpha_2|^2}
\int dZ\left[{\alpha_{12}^2
-4\alpha_1\alpha_2Z}\right]Z^{-S-2+s/12}(1-Z)^{S-s/4+2}\nonumber\\
&=&\frac{g^{2}}{|
\alpha_1\alpha_2|}\left[\frac{\alpha_{12}^2}{\alpha_2^2}
\frac{\Gamma(-S-1+s/12)}{\Gamma(2-s/6)}
-4\frac{\alpha_1}{\alpha_2}\frac{\Gamma(-S+s/12)}{\Gamma(3-s/6)}
\right]{\Gamma(S-s/4+3)}
\eea
In this case there is no value of $S$ for which there is convergence
at both $Z=0$ and $Z=1$, so we refrained from setting $s=24$ in
the powers as a regulator. When we attempt to set $s=24$, the
denominators of both terms blow up, suggesting that there is
no backward scattering in this process! Of course it
is a very special process, corresponding to maximal helicity violation.

Although the kinematics of these special 4 string processes 
do not correspond to fully elastic forward
scattering, since the
internal state of string 3 is not identical to that of string 2,
it does reflect the high energy behavior of forward scattering 
argued for in the next section for truly elastic scattering
process. 
\section{High energy four string scattering}
There is a general argument, based on the lightcone worldsheet description
\cite{mueller}, that the 
the forward elastic open string scattering probability
amplitude\footnote{In Lorentz covariant theories the probability
amplitude is obtained by dividing the covariant Feynman
amplitude by $\prod_k\sqrt{2p_k^+}$.} goes to a constant at high energies. 
In this argument high energy scattering is reached
by taking $p_1^+\to\infty$ at fixed $p_2^+$ and fixed
$p_2^++p_3^+=-p_1^+-p_4^+$. It is also assumed that the 
internal state of string 1 is
identical to the internal state of string 4, so that the ``large'' string
is elastic. Then $(p_1+p_2)^-\sim m^2/p^+_2$ stays finite. When $p^+_1
\to\infty$ the
lightcone worldsheet becomes very large while the effect of the interaction
is limited to a region of size of order 1. Since the speed of sound is finite
the amplitude must approach a constant as $p_1^+\to\infty$. In a Lorentz
invariant theory, the amplitude is ${\cal M}/\sqrt{p_1^+p_2^+p_3^+p_4^+}$ where 
${\cal M}$ is the invariant Feynman amplitude. Since $S=-(p_1+p_2)^2
\sim m^2p_1^+/p_2^+$,
it follows that ${\cal M}\propto S$ at large $S$ corresponding to
a leading Regge trajectory of intercept 1. 

But this result is a consequence of the lightcone parameterization of
the worldsheet, whether or not Lorentz invariance is met.
It is instructive to check this for general $s$. 
For the Lorentz invariant bosonic open string 
scattering amplitude ($s=0$), the limit $-S=(p_1+p_2)^2
\to\infty$ is evaluated by changing variables $Z=e^{-u}$
\bea
{\cal M}&\propto&\int_0^\infty du e^{-u(-S-1)}(1-e^{-u})^{-t-2}\sim
(-S-1)^{t+1}\Gamma(-t-1),\qquad S\to -\infty.
\eea
where $-t=(p_2+p_3)^2$ is the momentum transfer ($=0$ in the forward direction).

When the Grassmann dimension $s>0$, 
the analysis is complicated by the dependence of
the integrand on $x_\pm$. The high energy behavior is still controlled by
$Z\sim1$, so consulting (\ref{limitexes}) we find, holding $\alpha_{23}
=-\alpha_{14}\neq0$,
\bea
x_+&\sim&-\frac{\alpha_1}{\alpha_4},\qquad x_-\sim1,\qquad
x_+-x_-\sim -\frac{1}{\alpha_1}\left[\frac{\alpha_1\alpha_{14}}{\alpha_4}\right]\\
1-x_+&\sim&\frac{1}{\alpha_1}\left[\frac{\alpha_1\alpha_{14}}{\alpha_4}\right],\qquad 
1-x_-\sim-(1-Z)\frac{\alpha_3}{\alpha_{14}}\sim-\frac{1}{\alpha_1}
\left[u\frac{\alpha_1\alpha_3}{\alpha_{14}}\right]\\
Z-x_+&\sim&\frac{1}{\alpha_1}\left[\frac{\alpha_1\alpha_{14}}{\alpha_4}\right],\qquad 
Z-x_-\sim(1-Z)\frac{\alpha_2}{\alpha_{14}}\sim\frac{1}{\alpha_1}
\left[u\frac{\alpha_1\alpha_2}{\alpha_{14}}\right]
\label{alpha23notzero}\eea
The limit $\alpha_{23}\to0$ is not uniform. Indeed
setting $\alpha_3=-\alpha_2$ from the beginning, we find
\bea
x_\pm&=&1+\frac{1}{2\alpha_1}\left[(\alpha_2-\alpha_1)(1-Z)\pm{\sqrt{4\alpha_1\alpha_2(1-Z)+(\alpha_1-\alpha_2)^2(1-Z)^2}}\right]\\
&\sim&1+\frac{1}{2\alpha_1}\left[-\alpha_1u\pm{\sqrt{4\alpha_1\alpha_2u
+\alpha_1^2u^2}}\right]
\eea
From which we conclude
\bea
1-x_\pm&\sim&\frac{1}{2\alpha_1}\left[\alpha_1u\mp{\sqrt{4\alpha_1\alpha_2u
+\alpha_1^2u^2}}\right]\\
Z-x_\pm&\sim&\frac{1}{2\alpha_1}\left[-\alpha_1u\mp{\sqrt{4\alpha_1\alpha_2u
+\alpha_1^2u^2}}\right]\\
x_+-x_-&\sim&\frac{1}{\alpha_1}\left[{\sqrt{4\alpha_1\alpha_2u
+\alpha_1^2u^2}}\right]
\label{alpha230}\eea
Since the important region of integration at high energy is 
$1-Z\approx u=O(S^{-1})$ and
we are keeping $\alpha_1/S$ fixed, the factors in square brackets
are of order 1, for both cases, 
in the dominant integration region. We see that in this region
$x_\pm \sim O(1)$ and $1-Z$, $1-x_\pm$, $Z-x_\pm$, and $x_+-x_-$ all
scale as $\alpha_1^{-1}$. Now referring back to (\ref{openintegrand})
for the case $N=4$, we read off the total power of $\alpha_1^{-1}$:
\bea
&&{2{\bfs\gamma}_1\cdot{\bfs\gamma}_2
-s/24+2P_2\cdot P_3-s/24+2({\bfs\pi}_2+{\bfs\pi}_3)\cdot({\bfs\gamma}_1
+{\bfs\gamma}_2)+s/6}\nonumber\\
&&\quad =({\bfs\gamma}_1
+{\bfs\gamma}_2)^2-\frac{s}{8}+(P_2+P_3)^2-2\left(1-\frac{s}{48}\right)
+2({\bfs\pi}_2+{\bfs\pi}_3)\cdot({\bfs\gamma}_1
+{\bfs\gamma}_2)+\frac{s}{12}\nonumber\\
&&\quad =({\bfs\gamma}_1
+{\bfs\gamma}_2)^2+(P_2+P_3)^2-2
+2({\bfs\pi}_2+{\bfs\pi}_3)\cdot({\bfs\gamma}_1
+{\bfs\gamma}_2)\nonumber\\
&&\quad =(p_2+p_3)^2-2
+({\bfs\pi}_1+{\bfs\pi}_4)^2\eea
We conclude that the high energy behavior of the scattering amplitude is
\bea
A_4&\sim& \alpha_1^{1+t-({\bfs\pi}_1+{\bfs\pi}_4)^2},\qquad t\equiv-(p_2+p_3)^2
\eea
in accord with the general argument for $t=0$ provided that ${\bfs\pi}_4=-{\bfs\pi}_1$,
i.e. provided that the internal states of the ``long'' strings are identical.
 
We close with a simple example of a fully elastic scattering amplitude. 
We choose ${\bfs\pi}_4=-{\bfs\pi}_1$ {\it and}
${\bfs\pi}_3=-{\bfs\pi}_2$. Necessarily then we must have ${\bfs\gamma}_2
=-{\bfs\gamma}_1$. Let ${\bfs\gamma}$ be the $s/2$ vector with each component
equal to $\gamma=1/\sqrt{8}$ for the open string. Then a simple example
of elastic scattering would be
\bea
{\bfs\pi}_1={\bfs\pi}_2={\bfs\gamma},\qquad {\bfs\pi}_3={\bfs\pi}_4
=-{\bfs\gamma},\qquad {\bfs\gamma}_1=-{\bfs\gamma}_2=\pm{\bfs\gamma}
\eea
Then the integrand of the scattering amplitude formula 
becomes
\bea
&&\hskip-.5cm|Z|^{-S-2+s/4}|1-Z|^{-t-2}|x_1-x_2|^{-s/6}\nonumber\\
&&\times\begin{cases}|x_1|^{s/6}|x_2|^{-s/12}|1-x_1|^{-s/12}
|1-x_2|^{s/6}
|Z-x_1|^{s/6}|Z-x_2|^{-s/12}&{\rm for}\quad {\bfs\gamma}_1={\bfs\gamma}\\
|x_1|^{-s/12}|x_2|^{s/6}|1-x_1|^{s/6}|1-x_2|^{-s/12}
|Z-x_1|^{-s/12}|Z-x_2|^{s/6}&{\rm for}\quad 
 {\bfs\gamma}_1=-{\bfs\gamma}\end{cases}
\eea
The second case just interchanges $x_1\leftrightarrow x_2$. 

We first evaluate the
high energy limit with $\alpha_{23}\neq0$, changing integration variables to $Z=e^{-v/\alpha_1}$ and treating
$v=O(1)$ to get
\bea
{\cal M}
&\sim&\alpha_1^{1+t}\int_0^\infty dve^{-(-S-1+s/4)v/\alpha_1}
v^{-t-2+s/12}\left|{\alpha_{14}}\right|^{-s/6}
\nonumber\\
&&\times\begin{cases}\left|\frac{\alpha_1}{\alpha_4}\right|^{s/12}
\left|{\alpha_3}\right|^{s/6}
\left|{\alpha_2}\right|^{-s/12}&\qquad{\rm for}\quad {\bfs\gamma}_1={\bfs\gamma}\\
\left|\frac{\alpha_1}{\alpha_4}\right|^{-s/6}\left|{\alpha_3}\right|^{-s/12}
\left|{\alpha_2}\right|^{s/6}&\qquad{\rm for}\quad 
 {\bfs\gamma}_1=-{\bfs\gamma}\end{cases}\nonumber\\
&\sim&\alpha_1^{s/12}\left({-S}\right)^{t+1-s/12}
\Gamma(-t-1+s/12)\left|{\alpha_{14}}\right|^{-s/6}
\nonumber\\
&&\times\begin{cases}\left|\frac{\alpha_1}{\alpha_4}\right|^{s/12}
\left|{\alpha_3}\right|^{s/6}
\left|{\alpha_2}\right|^{-s/12}&\qquad{\rm for}\quad {\bfs\gamma}_1={\bfs\gamma}\\
\left|\frac{\alpha_1}{\alpha_4}\right|^{-s/6}\left|{\alpha_3}\right|^{-s/12}
\left|{\alpha_2}\right|^{s/6}&\qquad{\rm for}\quad 
 {\bfs\gamma}_1=-{\bfs\gamma}\end{cases}
\eea
Here we see that, with $\alpha_{23}\neq0$ and fixed, the coefficient of 
$\alpha_1^{t+1}$ has poles at $t=n+s/12-1$ which are the mass squared  
eigenvalues of the open protostring. The linear high energy behavior
at $t=0$ is the product of $(-S)^{1-s/12}$ and $\alpha_1^{s/12}$
netting precisely linear growth in the forward direction. 

Contrast this with the high energy limit taken with $\alpha_{23}=0$ from
the beginning:
\bea
{\cal M}&\sim& \alpha_1^{1+t}\int_0^\infty dv\ e^{-(-S-1+s/4)v/\alpha_1}
v^{-t-2}\left[{{4\alpha_2v
+v^2}}\right]^{-s/12}\nonumber\\
&&\times\left|\frac{1}{2}\left[v-{\sqrt{4\alpha_2v
+v^2}}\right]\right|^{-s/12} \left|\frac{1}{2}\left[v+{\sqrt{4\alpha_2v
+v^2}}\right]\right|^{s/6}\nonumber\\
&&\times\left|\frac{1}{2}\left[-v-{\sqrt{4\alpha_2v
+v^2}}\right]\right|^{s/6} \left|\frac{1}{2}\left[-v+{\sqrt{4\alpha_2v
+v^2}}\right]\right|^{-s/12}\nonumber\\
&\sim& \alpha_1^{1+t}\int_0^\infty dv\ e^{-(-S-1+s/4)v/\alpha_1}
v^{-t-2}\left[{{4
+v/\alpha_2}}\right]^{-s/12}\left|\sqrt{1+\frac{v}{4\alpha_2}}
+\sqrt{\frac{v}{4\alpha_2}}\right|^{s/2}\eea
We stress that the limit taken here is $\alpha_1, -S\to\infty$
at fixed ratio. The coefficient of the Regge behavior is a function
of $t$. Its pole locations are {\it not} those of the
particles of the theory: they correspond to a linear Regge trajectory of
intercept 1. Because the formula was obtained assuming $\alpha_3=-\alpha_2$,
the high energy behavior comes from the collision of two interaction
points on the lightcone worldsheet, and not from the long time propagation
of a protostring mass eigenstate as in the first limiting
procedure. The mismatch can be allowed because the Lorentz boost symmetry
generated by $M^{-k}$ is absent: for the protostring because
there is no transverse space and for $0<s<24$ because this part of the
Lorentz symmetry is broken. For the bosonic string ($s=0$), of course, 
there is no such mismatch.
\section{Concluding Remarks}
This article is devoted to the calculation of scattering amplitudes
for the protostring and a simple generalization
thereof. The three string amplitude with the strings
in arbitrary excited states was calculated in \cite{thornprotobits},
where the model was initially proposed. Here the focus has been
on general $N$ string amplitudes, but with the strings in states
with no oscillator excitations. These amplitudes are analogous
to the $N$ tachyon amplitudes of the bosonic string.

The scattering amplitudes are presented as integrals over the
Koba-Nielsen variables $Z_k$. The integrand includes factors
$|Z_k-Z_l|$ raised to momentum dependent powers, familiar
from the bosonic string. But in addition their are factors
$|x_r-Z_k|$ and $|x_r-x_s|$ raised to powers dependent on the
compactified momentum representing the Grassmann degrees of
freedom. Here the $x_r(Z)$ are the locations in the $z$-plane
of the break/join points of the lightcone worldsheet. If
$\rho(z)$ is the conformal 
map from the $z$ plane to the lightcone worldsheet, the $x_r$
are determined by the order $N-2$ polynomial equation
$d\rho/dz=0$. The presence of these other factors complicates
the singularity structure of the integrands.  

We studied in detail some simple
special cases. We found significant simplifications for
the maximal helicity (compactified momentum) violating $N$
string amplitudes. But factors involving $|x_r-x_s|$
remain, which become quite unwieldy for $N>4$. 
On the other hand four string amplitudes are 
manageable, because the $x$'s solve a quadratic equation,
$(x_1-x_2)^2$ being the discriminant. For the protostring case
$s=24$ this contribution is just a quadratic polynomial in $Z$.
So the four point functions are sums of Euler beta functions.
For forward open protostring scattering ($t=0$), the $S,t$ amplitude\footnote{
With $S=-(p_1+p_2)^2$ and $t=-(p_2+p_3)^2$, the $S,t$ amplitude
is the one with cyclic ordering $12341$. It has
poles in the $S$ and $t$ channels.} is just a 
sum of a finite number of poles at $S=1,2,3$. Curiously, the
$S,t$ amplitude for backward scattering ($t=4-S$) seems to vanish\footnote{The
argument for vanishing backward scattering relies on an analytic continuation
which may be suspect.} for the protostring.
This vanishing of backward scattering may
be specific to helicity violating amplitudes. It would be interesting
to explore whether other protostring amplitudes vanish.

We 
analyzed the high energy limit of selected 4 open string amplitudes,
in both helicity conserving and helicity violating processes. 
We confirmed that at large
energy in the forward direction the elastic amplitude is constant
at high energy, in line with the general argument in
\cite{mueller}. 
In a Lorentz covariant string theory in three
or more dimensions such behavior would signal a leading Regge trajectory
of unit intercept, which would imply massless gauge
particles in the theory. The models studied here are 
non-covariant for $0<s<24$, depending not only on the Lorentz invariants
$p_k\cdot p_l$ but also on the $+$ momentum components $p_k^+$.
Instead of the covariant 
Regge behavior $(p_1\cdot p_2)^{t+1}/\sqrt{p_1^+p_2^+p_3^+p_4^+}$
we get $(p_1^{+})^{s/12} (p_1\cdot p_2)^{t+1-s/12}/\sqrt{p_1^+p_2^+p_3^+p_4^+}$. 
Here the Regge trajectory $\alpha(t)=t+1-s/12$ reflects the
spectrum of the string models of the present paper: for 
the protostring ($s=24$) it is $t-1$, implying a mass gap! 
The constant high energy behavior when $t=0$ generally applies \cite{mueller} 
when $(p_1\cdot p_2)/p_1^+$ is fixed as $p_1\cdot p_2\to\infty$.
This is how constant high energy behavior is consistent with a Regge
trajectory intercept less than unity. 

There is a lot of work still to be done on these models. The amplitudes
involving more than 4 strings have been obtained, but their physical
properties remain to be investigated. One should also be able
to calculate these amplitudes in the original unbosonized language
and compare results to those of the present paper. This
comparison
should clarify issues of Fermi statistics, which the process
of bosonization has obscured.

The protostring ($s=24$) moves only in 1 space dimension, so the
entire Lorentz group is $O(1,1)$, and this symmetry is maintained
in the construction. As we have said the protostring is predicted by
the simplest of string bit models. In the string limit, 
the Grassmann dimension $s$
effectively interpolates between the bosonic string $s=0$ and the
protostring ($s=24$). We have noted that the protostring has
the worldsheet field content of the superstring. This suggests that
besides being an interesting system in its own right, it may also
be a stepping stone to a solid foundation of superstring theory. 
In its present form, the 8 ``transverse coordinates'' one obtains
by bosonizing 16 of the protostring's Grassmann dimensions are
all compactified in a way such that the KK momentum is quantized in half
odd integers. Perhaps there is a way to tweak the Hamiltonian of
the string bit model underlying the protostring 
to shift this quantization to integers. An additional tweak would
be needed to provide a large compactification radius for at least
two of these emergent transverse coordinates. We leave the
pursuit of these goals to future research.
\vskip18pt
\noindent\underline{Acknowledgments}: I would like to thank Gaoli Chen
and Songge Sun for helpful comments on the manuscript.
This research was supported in part by the Department
of Energy under Grant No. DE-SC0010296.

\appendix
\section{Determinant
 for the Lightcone Worldsheet}
\subsection{Open String}
In this appendix we review Mandelstam's calculation of the
determinant and Jacobian factors for the bosonic string
\cite{mandelstamdet}.
The quantities $Z_k$, with $k=1\cdots (N-1)$, and $x_r$, 
with $r=1\cdots(N-2)$ are determined by the map from the
upper-half Koba-Nielsen plane ($z$) to the lightcone world sheet
($\rho=\tau+i\sigma$):
\bea
\rho&=&\sum_{k=1}^{N-1}\alpha_k\ln(z-Z_k),\qquad {d\rho\over dz}\bigg|_{z=x_r}=0\label{rhoofzo}\\
{d\rho\over dz}&=&\sum_{k=1}^{N-1}{\alpha_k\over z-Z_k}
={\sum_{k=1}^{N-1}\alpha_k\prod_{l\neq k}(z-Z_l)\over
\prod_k(z-Z_k)}=-\alpha_N{\prod_r (z-x_r)\over\prod_k(z-Z_k)}\label{rhoprimeo}\\
{d^2\rho\over dz^2}\bigg|_{z=x_s}&=&-\alpha_N{\prod_{r\neq s} (x_s-x_r)\over\prod_k(x_s-Z_k)}
\label{rho2primeo}
\eea
where the last line is true because the factor $(z-x_s)$ in the numerator
must be killed by the derivative to get a nonzero contribution.
The asymptotic strings at $\tau=\pm\infty$ are mapped from
the $Z_k$. In this notation $Z_N=\infty, Z_1=0$. 

We next determine the worldsheet determinant. We do this
by executing a conformal
transformation from the Koba-Nielsen plane to the lightcone worldsheet.
We shall need 
\bea
\Sigma\equiv\ln\left|\frac{d\rho}{dz}\right|
=\ln|\alpha_N|-\sum_{k=1}^{N-1}\ln|z-Z_k|+\sum_{r=1}^{N-2}\ln|z-x_r|
\label{sigmawsopen}\eea
Clearly $\partial_y\Sigma=0$ on the real axis. Since the points
$z=Z_k,x_s$ are singular, we deform the boundary near those points into small
semicircles, in the upper half plane, of radii $\epsilon_k,\epsilon_r$ 
respectively. The radius
$\epsilon_k$ near $Z_k$ can be interpreted in terms of a large time
$T_k$ for the asymptotic string $k$. From the mapping function we find
\bea
\epsilon_k=e^{T_k/\alpha_k}\prod_{l\neq k}|Z_l-Z_k|^{-\alpha_l/\alpha_k}
\label{epsilontopen}
\eea
The string $N$ is asymptotic at large $z$. If $R$ is the radius of
a large semi-circle, we have from the mapping function
\bea
T_N\sim -\alpha_N\ln R, \qquad R\sim e^{-T_N/\alpha_N}.
\eea
On the other hand the radius $\epsilon_s$ near $x_s$ is a temporary regulator,
which maps onto a circular deformation of the boundary
near the corresponding interaction
point on the lightcone worldsheet. From the mapping function we see that
the radius of this regulating circle on the worldsheet is given
by
\bea
\delta_s&=&{1\over2}\epsilon_s^2\bigg|{d^2\rho\over dz^2}\bigg|_{z=x_s}
={1\over2}\epsilon_s^2|\alpha_N|{\prod_{r\neq s} 
|x_s-x_r|\over\prod_k|x_s-Z_k|}
\\
\epsilon_s&=&\sqrt{2\delta_s\over|\alpha_N|}
{\prod_k|x_s-Z_k|^{1/2}\over\prod_{r\neq s} 
|x_s-x_r|^{1/2}}\\ 
\prod_s\epsilon_s&=&|\alpha_N|^{-N+3/2}\prod_k|\alpha_k|^{1/2}
\prod_s\sqrt{2\delta_s}
{\prod_{l\neq k}|Z_l-Z_k|^{1/2}\over\prod_{r\neq s} 
|x_s-x_r|^{1/2}}
\eea
To calculate the determinant for the lightcone worldsheet, we start
with the determinant for the region in the upper-half $z$-plane bounded by the
real axis, the large radius $R$ semi-circle, and the small radius
$\epsilon_k$ semi-circles. Then we apply the
generalized Kac-McKean-Singer formula \cite{kacdrum,mckeans,thornwsdet} 
to transform to the determinant for the worldsheet.

In this case, the boundary conditions are either
Dirichlet everywhere or Neumann everywhere.
Then in the limit of large $R$
and small $\epsilon$, factorization implies that the $z$-plane
figure determinant has the behavior
\bea
-{1\over2}\Tr\ln(-\nabla^2)_z&\sim& {5\over24}\ln R +{1\over24}\sum_k
\ln\epsilon_k +{\rm const}
\eea
where the constant term, representing the determinant for the
upper half plane with the same boundary conditions everywhere,
has nothing to depend on! 

Next we develop the transformation of the determinant from this
$z$-plane figure to the lightcone worldsheet using (\ref{sigmawsopen}).
Clearly $\partial_y\Sigma=-\partial_n\Sigma=0$ on the real axis. Thus
the change formula receives contributions from the corners
and semi-circles only. For $z$ near $Z_k$ put $z=Z_k+r e^{i\varphi}$
and approximate
\bea
\Sigma\approx\ln|\alpha_N|-\ln r-\sum_{l\neq k}^{N-1}\ln|Z_l-Z_k|
+\sum_{r=1}^{N-2}\ln|Z_k-x_r|,\qquad \partial_n\Sigma\approx {1\over r}
\eea
Then
\bea
\Delta_{\epsilon_k}&=&\left[{1\over24}-{1\over12}+{1\over8}\right]\Sigma
={1\over12}\ln\left({|\alpha_N|\over\epsilon_k}{\prod_r|Z_k-x_r|
\over\prod_{l\neq k}|Z_k-Z_l|}\right)=
\ln\left({|\alpha_k|\over\epsilon_k}\right)^{1/12}
\label{changeNNopen}
\eea
The three terms in square brackets are the $\int dl\Sigma\partial_n\Sigma$
term the extrinsic curvature term (negative here) 
and the two corners at this semi-circle respectively.

For $z=x_s+re^{i\varphi}$, on the other hand we have
\bea
\Sigma\approx\ln|\alpha_N|+\ln r-\sum_{l}^{N-1}\ln|Z_l-x_s|
+\sum_{r\neq s}\ln|x_s-x_r|,\qquad \partial_n\Sigma\approx -{1\over r}
\eea
Then
\bea
\Delta_{\epsilon_s}=\left[-{1\over24}\right]\Sigma=-\frac{1}{24}\ln\frac{2\delta_s}{\epsilon_s}
\label{changexopen}
\eea
Note that in this case only the $\int\Sigma\partial_n\Sigma$ term contributes
since there is no singularity in the initial surface at $z=x_s$: the singularity
comes only in $\Sigma$ which is determined by the mapping function.

Finally for the large semi-circle, $\Sigma\approx -\ln (r/|\alpha_N|)$,
$\partial_n\Sigma\approx -1/r$, and
\bea
\Delta_R=\left[-{1\over24}+{1\over12}+{1\over8}\right]\Sigma
=-{1\over6}\ln {R\over|\alpha_N|}
\eea
Combining all the contributions,
we have
\bea
{\det}^{-1/2}(-\nabla^2)_\rho&=&{\det}^{-1/2}(-\nabla^2)_z
\left({|\alpha_N|\over R}\right)^{1/6}\prod_k\left({|\alpha_k|\over\epsilon_k}
\right)^{1/12}\prod_r\left(\frac{\epsilon_r}{2\delta_r}\right)^{1/24}\nonumber\\
&=&C{|\alpha_N|}^{1/6}\exp\left\{-\sum_{k=1}^N{T_k\over24\alpha_k}\right\}
\prod_{k\neq l}|Z_k-Z_l|^{\alpha_k/24\alpha_l}
\prod_k{|\alpha_k|}^{1/12}\nonumber\\
&&\left[{\prod_r(2\delta_r)\over|\alpha_N|^{N-2}}
{\prod_k\prod_r|x_r-Z_k|\over\prod_{r\neq s} 
|x_s-x_r|}\right]^{1/48}\prod_r(2\delta_r)^{-1/24}\nonumber\\
&=&C\prod_r(2\delta_r)^{-1/48}
\prod_{k=1}^N{|\alpha_k|^{1/8}\over|\alpha_k|^{1/48}}
\prod_{k\neq l}|Z_k-Z_l|^{\alpha_k/24\alpha_l}
\nonumber\\
&&\left[{1\over|\alpha_N|^{N-3}}
{\prod_{k< l}|Z_l-Z_k|\over\prod_{r< s} 
|x_s-x_r|}\right]^{1/24}\exp\left\{-\sum_{k=1}^N{T_k\over24\alpha_k}\right\}
\label{opendeterminant}
\eea
If there are $n_b$ bosonic worldsheet dimensions, this entire factor should be 
raised to the power $n_b$.

The worldsheet path integral is this determinant factor times a factor
$e^{iW_c}$ which arises from removing boundary data in the path integral
by shifting the ${\bfs x}$ by the classical solution that satisfies those
boundary data as shown in the text. 
Among other things $e^{iW_c}$ includes factors 
$R^{-{\bfs p}^2}\prod_k\epsilon_k^{{\bfs p}^2}$ in the limit 
that the $-T_k/\alpha_k$ get large.
If $W_c=\sum_{kl}p_k N(\rho_k,\rho_l)p_l/2$ is 
expressed in terms of a Neumann function, these factors arise from
the diagonal $l=k$ terms. The rest of these diagonal terms, 
combined with the factors $|\alpha_k|^{1/8}$, provide a factor of
the ground string wave function for each external string. 
The $N$ ground string scattering amplitude is
obtained by amputating these ground state wave functions
together with the factors $e^{\sum_k({\bfs p}_k^2-d/24)T_k/\alpha_k}
=e^{\sum_k T_kP^-_k}$
from the path integral and integrating over the interaction times
$\int d\tau_1\cdots d\tau_{N-2}$
where $\rho_r=\tau_r+i\sigma_r$ are the locations of the $N-2$ interaction
points on the worldsheet. By translational invariance in $x^+$
the integrand after amputation will acquire a factor
$e^{a\sum_kP^-_k}$ if all the $\tau_r$ are translated by $a$. This means
that integrating over one of the $\tau_r$ simply produces
a $P^-$ conserving delta function. The coefficient of this delta function
is just the integral over only $N-3$ of the $\tau_r$. Note that
$\sum_k\alpha_k=0$ by the lightcone worldsheet construction and
$\sum_k P_k=0$ when Neumann conditions are chosen for the                
${\bfs x}$ integrals as explained in Section 3 .       
\bea
{\cal A}&=&\int d\tau_2\cdots d\tau_{N-2}\left[{\det}^{-n_b/2}(-\nabla^2)_\rho e^{iW_c}\right]_{\rm amputated}
\eea
where we have set $\tau_1=0$ and understand that $\sum_kP^-_k=0$.

The final result for $[e^{iW_c}]_{\rm amputated}$ includes the off diagonal
terms in its Neumann function representation, together with the 
parts of $\epsilon_k$ that remain after amputating $e^{\sum_k T_kP^-_k}$:
For the ordinary bosonic coordinates the result is
\bea
\left[e^{iW_c}\right]_{\rm amputated}&=&\prod_{k<l}|Z_l-Z_k|^{2{\bfs p}_k\cdot
{\bfs p}_l}\left(\prod_{k\neq l}|Z_k-Z_l|
\right)^{-\alpha_l{\bfs p}_k^2/\alpha_k}
\eea
The amputated determinant drops the exponential dependence on $T_k$
and the factor $\prod_k|\alpha_k|^{n_b/8}$:
\bea
\left[{\det}^{-n_b/2}(-\nabla^2)_\rho\right]_{\rm amputated}
&=&C\prod_r(2\delta_r)^{-n_b/48}\prod_{k=1}^N{1\over|\alpha_k|^{n_b/48}}
\prod_{k\neq l}|Z_k-Z_l|^{n_b\alpha_k/24\alpha_l}
\nonumber\\
&&
\qquad\qquad\left[\frac{1}{|\alpha_N|^{N-3}}
{\prod_{k< l}|Z_l-Z_k|\over\prod_{r< s} 
|x_s-x_r|}\right]^{n_b/24}.
\label{knmeasure}
\eea
Notice that when the two are combined the net power of $|Z_k-Z_l|$
simplifies nicely
$$p_k\cdot p_l={\bfs p}_k\cdot {\bfs p}_l
-p^+_kp^-_l-p^-_kp^+_l={\bfs p}_k\cdot {\bfs p}_l
-\alpha_k({\bfs p}^2_l-n_b/24)/2\alpha_l
-\alpha_l({\bfs p}^2_k-n_b/24)/2\alpha_k.$$
It is convenient to change integration variables from the $\tau$'s to the
$Z$'s. Mandelstam's result for the Jacobian is (taking $Z_1,Z_{N-1},Z_N=0,1,
\infty$ respectively)
\bea
{\cal J}={\partial(\tau_2,\cdots,\tau_{N-2})\over
\partial(Z_2,\cdots,Z_{N-2})}&=&\left[{1
\over|\alpha_N|^{N-3}}
{\prod_{k< l}|Z_l-Z_k|\over\prod_{r< s} 
|x_s-x_r|}\right]^{-1},\\
{\cal J}\left[{\det}^{-n_b/2}(-\nabla^2)_\rho\right]_{\rm amputated}
&=&C\prod_r(2\delta_r)^{-n_b/48}\prod_{k=1}^N{1\over|\alpha_k|^{n_b/48}}
\prod_{k\neq l}|Z_k-Z_l|^{n_b\alpha_k/24\alpha_l}\nonumber\\&&
\qquad\qquad\left[\frac{1}{|\alpha_N|^{N-3}}
{\prod_{k< l}|Z_l-Z_k|\over\prod_{r< s} 
|x_s-x_r|}\right]^{(n_b-24)/24}
\eea
so that the scattering amplitude for the purely bosonic string ($n_b=d$) 
becomes
\bea
{\cal A}&=&C\prod_r(2\delta_r)^{-n_b/48}\prod_{k=1}^N{1\over|\alpha_k|^{n_b/48}}
\int dZ_2\cdots dZ_{N-2}
\prod_{k<l}|Z_k-Z_l|^{2p_k\cdot p_l}
\nonumber\\
&&
\qquad\qquad\left[{1\over|\alpha_N|^{N-3}}
{\prod_{k< l}|Z_l-Z_k|\over\prod_{r< s} 
|x_s-x_r|}\right]^{(d-24)/24}
\eea
The factor raised to the power $d-24$ depends on the Lorentz frames
so the critical dimension $D=26$ is necessary for Lorentz invariance
\cite{goddardgrt}, in which case ${\cal A}$ is proportional
to the $N$ particle dual resonance amplitude.
Of course factorization implies that $C=g^{N-2}$
and $\delta_r=\delta$, independent of $r$. Then 
$\prod_r(2\delta_r)=(2\delta)^{N-2}$
so $\delta$ can be absorbed in the coupling constant.

\subsection{Closed String}
For the closed string the map from the
whole Koba-Nielsen plane ($z$) to the lightcone world sheet
($\rho=\tau+i\sigma$) is nearly identical to that for the open string.
\bea
\rho&=&\frac{1}{2}\sum_{k=1}^{N-1}\alpha_k\ln(z-Z_k),\qquad {d\rho\over dz}\bigg|_{z=x_r}=0\label{rhoofzc}\\
{d\rho\over dz}&=&\frac{1}{2}\sum_{k=1}^{N-1}{\alpha_k\over z-Z_k}
={\sum_{k=1}^{N-1}\alpha_k\prod_{l\neq k}(z-Z_l)\over
2\prod_k(z-Z_k)}=-\frac{\alpha_N}{2}{\prod_r (z-x_r)\over\prod_k(z-Z_k)}
\label{rhoprimec}\\
{d^2\rho\over dz^2}\bigg|_{z=x_s}&=&
-\frac{\alpha_N}{2}{\prod_{r\neq s} (x_s-x_r)\over\prod_k(x_s-Z_k)}
\label{rho2primec}
\eea
The quantities $Z_k$, with $k=1\cdots (N-1)$, and $x_r$, 
with $r=1\cdots(N-2)$ can now be anywhere in the complex plane.
The factors of $1/2$ on the right are to normalize the range of
$\sigma$ on string $k$ to $\pi\alpha_k$, since the phase of $z-Z_k$
advances by $2\pi$ as $z$ encircles $Z_k$.
The asymptotic strings at $\tau=\pm\infty$ are mapped from
the $Z_k$. In this notation $Z_N=\infty, Z_1=0$. 

We next turn to the transformation of the determinant.
\bea
\Sigma=\ln\left|\frac{\alpha_N}{2}\right|
-\sum_{k=1}^{N-1}\ln|z-Z_k|+\sum_{r=1}^{N-2}\ln|z-x_r|
\label{closedsigmaws}
\eea
Since the points
$z=Z_k,x_s$ are singular, we cut out small circular disk  
of radii $\epsilon_k,\epsilon_r$ about each of those points.
respectively. The radius
$\epsilon_k$ near $Z_k$ can be interpreted in terms of a large time
$T_k$ for the asymptotic string $k$. From the mapping function we find
\bea
\epsilon_k=e^{2T_k/\alpha_k}\prod_{l\neq k}|Z_l-Z_k|^{-\alpha_l/\alpha_k}
\label{epsilont}
\eea
The string $N$ is asymptotic at large $z$, so we cut out the region
$|Z|>R$. We take $R$ large,  and referring to the mapping function
\bea
T_N\sim -\frac{\alpha_N}{2}\ln R, \qquad R\sim e^{-2T_N/\alpha_N}.
\eea
On the other hand the radius $\epsilon_s$ near $x_s$ is a temporary regulator,
which maps onto a circular deformation of the boundary
near the corresponding interaction
point on the lightcone worldsheet. From the mapping function we see that
the radius of this regulating circle on the worldsheet is given
by
\bea
\delta_s&=&{1\over2}\epsilon_s^2\bigg|{d^2\rho\over dz^2}\bigg|_{z=x_s}
={1\over4}\epsilon_s^2|\alpha_N|{\prod_{r\neq s} 
|x_s-x_r|\over\prod_k|x_s-Z_k|}
\\
\epsilon_s&=&\sqrt{4\delta_s\over|\alpha_N|}
{\prod_k|x_s-Z_k|^{1/2}\over\prod_{r\neq s} 
|x_s-x_r|^{1/2}}\\ 
\prod_s\epsilon_s&=&|\alpha_N|^{-N+3/2}\prod_k|\alpha_k|^{1/2}
\prod_s\sqrt{4\delta_s}
{\prod_{l\neq k}|Z_l-Z_k|^{1/2}\over\prod_{r\neq s} 
|x_s-x_r|^{1/2}}
\eea
To calculate the determinant for the lightcone worldsheet, we start
with the determinant for the region in the $z$-plane with disks about
the $Z_k$ removed and bounded by the large radius $R$ circle, 
 Then we apply the
generalized Kac-McKean-Singer formula \cite{kacdrum,mckeans,thornwsdet} 
to transform to the determinant for the worldsheet.

We take boundary conditions on the circles to be Dirichlet
or Neumann. Then in the limit of large $R$
and small $\epsilon$, factorization implies that the $z$-plane
figure determinant has the behavior
\bea
-{1\over2}\Tr\ln(-\nabla^2)_z&\sim& {1\over6}\ln R -{1\over6}\sum_k
\ln\epsilon_k+{\rm const}
\eea
where the constant term, representing the determinant for the
whole plane with the same boundary conditions everywhere,
has nothing to depend on! 

Next we develop the transformation of the determinant from this
$z$-plane figure to the lightcone worldsheet:
The change formula receives contributions from the circular
boundaries only. For $z$ near $Z_k$ put $z=Z_k+r e^{i\varphi}$
and approximate
\bea
\Sigma\approx\ln|\alpha_N|-\ln r-\sum_{l\neq k}^{N-1}\ln|Z_l-Z_k|
+\sum_{r=1}^{N-2}\ln|Z_k-x_r|,\qquad \partial_n\Sigma\approx {1\over r}
\eea
Then
\bea
\Delta_{\epsilon_k}&=&\left[\frac{1}{12}-{1\over6}\right]\Sigma
=-{1\over12}\ln\left({|\alpha_N|\over\epsilon_k}{\prod_r|Z_k-x_r|
\over\prod_{l\neq k}|Z_k-Z_l|}\right)=
\ln\left({|\alpha_k|\over\epsilon_k}\right)^{-1/12}
\label{changeNN}
\eea
The terms in square brackets are the $\int dl\Sigma\partial_n\Sigma$
term and the extrinsic curvature term (negative here). 

For $z=x_s+re^{i\varphi}$, on the other hand we have
\bea
\Sigma\approx\ln|\alpha_N|+\ln r-\sum_{l}^{N-1}\ln|Z_l-x_s|
+\sum_{r\neq s}\ln|x_s-x_r|,\qquad \partial_n\Sigma\approx -{1\over r}
\eea
Then
\bea
\Delta_{\epsilon_s}=\left[-{1\over12}\right]\Sigma=
-\frac{1}{12}\ln\left(\epsilon_s|\alpha_N|
\frac{\prod_{r\neq s}|x_s-x_r|}{\prod_l|Z_l-x_s|}\right)
=-\frac{1}{12}\ln\frac{4\delta_s}{\epsilon_s}
\label{changex}
\eea
Finally for the large circle, $\Sigma\approx -\ln (r/|\alpha_N|)$,
$\partial_n\Sigma\approx -1/r$, and
\bea
\Delta_R=\left[-{1\over12}+{1\over6}\right]\Sigma=\frac{1}{12}\Sigma
=-{1\over12}\ln {R\over|\alpha_N|}
\eea
Combining all the contributions,
we have
\bea
{\det}^{-1/2}(-\nabla^2)_\rho&=&{\det}^{-1/2}(-\nabla^2)_z
\left({|\alpha_N|\over R}\right)^{1/12}\prod_k\left({|\alpha_k|\over\epsilon_k}
\right)^{-1/12}\prod_s\left({4\delta_s\over\epsilon_s}
\right)^{-1/12}\nonumber\\
&=&K{|\alpha_N|}^{1/12}R^{1/12}\prod_k\epsilon_k^{-1/12}\prod_r\epsilon_r^{1/12}
\prod_k{|\alpha_k|}^{-1/12}\prod_s(4\delta_s)^{-1/12}
\eea
We see that the $R$, $\epsilon_k$, $\epsilon_s$, and $\delta_s$ 
dependence of this
result is just the square of the corresponding dependence of the 
open string determinant with $\delta_s\to 2\delta_s$:
\bea
\frac{{\det}^{-1/2}(-\nabla^2)_{\rm closed}}{{\det}^{-1}(-\nabla^2)_{\rm open}
(\delta_s\to 2\delta_s)}
&=&\frac{K}{C^2}\frac{{|\alpha_N|}^{1/12}\prod_k{|\alpha_k|}^{-1/12}\prod_s(\delta_s)^{-1/12}}{|\alpha_N|^{1/3}\prod_k|\alpha_k|^{1/6}\prod_r(\delta_r)^{-1/12}}\nonumber\\
&=&\frac{K}{C^2}{|\alpha_N|}^{-1/4}\prod_{k<N}{|\alpha_k|}^{-1/4}
=\frac{K}{C^2}{\prod_{k\leq N}{|\alpha_k|}^{-1/4}}\nonumber
\eea
From this we read off from (\ref{opendeterminant}), 
sending $T_k,\delta_r\to 2T_k, 2\delta_r$
in the square of the open string determinant,
\bea
{\det}^{-1/2}(-\nabla^2)_\rho&=&K\prod_s(4\delta_s)^{-1/24}
\prod_{k=1}^N{1\over|\alpha_k|^{1/24}}
\prod_{k\neq l}|Z_k-Z_l|^{\alpha_k/12\alpha_l}
\nonumber\\
&&\left[{\frac{1}{|\alpha_N|^{N-3}}}
{\prod_{k< l}|Z_l-Z_k|\over\prod_{r< s} 
|x_s-x_r|}\right]^{1/12}\exp\left\{-\sum_{k=1}^N{T_k\over6\alpha_k}\right\}
\eea
If there are $n_b$ transverse bosonic dimensions this entire factor should be 
raised to the power $n_b$.

The construction of scattering amplitudes follows the same steps as
for the open string: one amputates external string wave functions
and exponential $T_k$ factors. In addition to integrating over interaction times one also integrates over the break/join location on the string, so the
amplitude takes the form:
\bea
{\cal A}&=&\int d\tau_2d\sigma_2\cdots d\tau_{N-2}d\sigma_{N-2}
\left[{\det}^{-n_b/2}(-\nabla^2)_\rho e^{iW_c}\right]_{\rm amputated}
\eea
where we have set $\tau_1=0$ and understand that $\sum_kP^-_k=0$.

For ordinary bosonic coordinates the relevant expressions are
\bea
\left[e^{iW_c}\right]_{\rm amputated}&=&\prod_{k<l}|Z_l-Z_k|^{{\bfs p}_k\cdot
{\bfs p}_l}\left(\prod_{k\neq l}|Z_k-Z_l|
\right)^{-\alpha_l{\bfs p}_k^2/(2\alpha_k)}\nonumber\\
\left[{\det}^{-n_b/2}(-\nabla^2)_\rho\right]_{\rm amputated}
&=&K\prod_r(4\delta_r)^{-n_b/24}\prod_{k=1}^N\frac{1}{|\alpha_k|^{n_b/24}}
\prod_{k\neq l}|Z_k-Z_l|^{n_b\alpha_k/12\alpha_l}
\nonumber\\
&&
\qquad\qquad\left[{1\over|\alpha_N|^{N-3}}
{\prod_{k< l}|Z_l-Z_k|\over\prod_{r< s} 
|x_s-x_r|}\right]^{n_b/12}
\eea
To change integration variables from the $\tau$'s to the
$Z$'s, Mandelstam's result for the Jacobian in the closed string
case \cite{mandelstamdet} is (taking $Z_1,Z_{N-1},Z_N=0,1,
\infty$ respectively)
\bea
|{\cal J}|^2=\left|{\partial(\rho_2,\cdots,\rho_{N-2})\over
\partial(Z_2,\cdots,Z_{N-2})}\right|^2&=&\left[{1
\over|\alpha_N|^{N-3}}
{\prod_{k< l}|Z_l-Z_k|\over\prod_{r< s} 
|x_s-x_r|}\right]^{-2},\\
|{\cal J}|^2\left[{\det}^{-n_b/2}(-\nabla^2)_\rho\right]_{\rm amputated}
&=&K\prod_r(4\delta_r)^{-n_b/24}\prod_{k=1}^N\frac{1}{|\alpha_k|^{n_b/24}}
\prod_{k\neq l}|Z_k-Z_l|^{n_b\alpha_k/12\alpha_l}
\nonumber\\
&&
\qquad\qquad\left[{1\over|\alpha_N|^{N-3}}
{\prod_{k< l}|Z_l-Z_k|\over\prod_{r< s} 
|x_s-x_r|}\right]^{(n_b-24)/12}
\eea
so that bosonic string scattering amplitude becomes
\bea
{\cal A}&=&K\prod_r(4\delta_r)^{-n_b/24}\prod_{k=1}^N{1\over|\alpha_k|^{n_b/24}}
\int d^2Z_2\cdots d^2Z_{N-2}
\prod_{k<l}|Z_k-Z_l|^{p_k\cdot p_l}
\nonumber\\&&
\qquad\qquad\left[{1\over|\alpha_N|^{N-3}}
{\prod_{k< l}|Z_l-Z_k|\over\prod_{r< s} 
|x_s-x_r|}\right]^{(n_b-24)/12}
\eea
The factor raised to the power $n_b-24$ depends on the Lorentz frames
so the critical dimension $n_b=24$ is necessary for Lorentz invariance
\cite{goddardgrt}.
\section{Alternative form for the measure}
Inspecting $d\rho/dz$ (see (\ref{rhoprimeo}),  (\ref{rhoprimec}))
we have the identity
\bea
-\alpha_N\prod_r (z-x_r)&=&\sum_{k=1}^{N-1}\alpha_k\prod_{l\neq k}(z-Z_l)\\
\eea
Putting $z=Z_m$ in this equation, we find
\bea
-\alpha_N\prod_r (Z_m-x_r)&=&\sum_{k=1}^{N-1}\alpha_k\prod_{l\neq k}(Z_m-Z_l)
=\alpha_m\prod_{l\neq m}(Z_m-Z_l)\label{identity1}\\
|\alpha_N|^{N}\prod_{m,r}|Z_m-x_r|&=&\prod_{m=1}^{N}|\alpha_m|
\prod_{l\neq k}|Z_k-Z_l|
\label{identity2}
\eea
Using this last equation,
the measures can be put in a form more useful for the protostring
scattering amplitudes:
\bea
&&\hskip-1.5cm\left[\left|{\partial T\over\partial Z}\right| 
{\det}^{-n_b/2}(-\nabla^2)\right]_{\rm open}=\prod_{k\neq l}|Z_k-Z_l|^{n_b\alpha_k/24\alpha_l}\nonumber\\
&&\times\prod_{k=1}^N{1\over \sqrt{|
\alpha_k|}}
\left[{\prod_{k<N}|\alpha_k|\over
|\alpha_N|}\right]^{(24-n_b)/16}\left[{\prod_{r<t}|x_t-x_r|
\prod_{m<l}|Z_l-Z_m|
\over \prod_{l,r}|Z_l-x_r|}\right]^{(24-n_b)/24},
\label{jacobdetopen2}
\eea
and
\bea
&&\hskip-1.5cm\left[\left|{\partial T\over\partial Z}\right|^2{\det}^{-n_b/2}(-\nabla^2)\right]_{\rm closed}
=\prod_{k\neq l}|Z_k-Z_l|^{n_b\alpha_k/12\alpha_l}\nonumber\\
&\times&\prod_{k=1}^N{1\over|
\alpha_k|}
\left[{\prod_{k<N}|\alpha_k|\over
|\alpha_N|}\right]^{(24-n_b)/8}\left[{\prod_{r<t}|x_t-x_r|
\prod_{m<l}|Z_l-Z_m|
\over \prod_{l,r}|Z_l-x_r|}\right]^{(24-n_b)/12}.
\label{jacobdetclosed2}
\eea

\end{document}